\documentclass[conference]{IEEEtran}
\usepackage{xspace}
\usepackage{subcaption}
\usepackage[table]{xcolor}
\usepackage{tabularx,booktabs}
\usepackage[frozencache]{minted}
\usepackage{adjustbox}
\usepackage{pgfplots}
\usepgfplotslibrary{statistics}
\usepackage[nomessages]{fp}
\usepackage{tikz,stackengine}
\usepackage{multicol}
\usepackage{makecell}
\usetikzlibrary{chains,arrows,matrix,decorations.markings,backgrounds,automata,shapes,snakes,positioning,calc}

\definecolor{Green}{rgb}{0.05, 0.5, 0.06}
\definecolor{Red}{rgb}{0.89, 0.26, 0.2}
\definecolor{darkcerulean}{rgb}{0.03, 0.27, 0.49}
\definecolor{shamrockgreen}{rgb}{0.0, 0.62, 0.38}
\definecolor{brass}{rgb}{0.71, 0.65, 0.26}
\definecolor{antiquefuchsia}{rgb}{0.57, 0.36, 0.51}
\newcommand{\vary}[1]{{\color{darkcerulean} $\mathtt{y_{#1}}$}}
\newcommand{\varz}[1]{{\color{shamrockgreen}$\mathtt{z_{#1}}$}}
\newcommand{\varu}[1]{{\color{brass}$\mathtt{u_{#1}}$}}
\newcommand{\varyp}[1]{{\color{antiquefuchsia}$\mathtt{y'_{#1}}$}}
\newcommand{\dream}{\texttt{DREAM++}\xspace}

\DeclareRobustCommand\single{\tikz \node[draw, very thick, circle, inner sep=0.08cm] at (0,0) {};}

\definecolor{colorg1}{RGB}{217,240,163}
\definecolor{colorg2}{RGB}{173,221,142}
\definecolor{colorg3}{RGB}{120,198,121}
\definecolor{colorg4}{RGB}{49,163,84}
\definecolor{colorg5}{RGB}{0,104,55}
\definecolor{color1}{RGB}{35,110,150}
\definecolor{color2}{RGB}{21,178,211}
\definecolor{color3}{RGB}{243,135,47}
\definecolor{color4}{RGB}{255,215,0}
\definecolor{color5}{RGB}{255,59,143}

\definecolor{my-bg}{rgb}{0.95,0.95,0.95}

\makeatletter
\tikzstyle{chart}=[
legend label/.style={font={},anchor=west,align=left},
legend box/.style={rectangle, draw, minimum size=8pt},
axis/.style={black,semithick,->},
axis label/.style={anchor=east,font={}},
]
\tikzstyle{bar chart}=[
chart,
bar width/.code={
    \pgfmathparse{##1/2}
    \global\let\bar@w\pgfmathresult
},
bar/.style={very thick, draw=white},
bar label/.style={font={\bf},anchor=north},
bar value/.style={font={}},
bar width=.75,
]
\tikzstyle{pie chart}=[
chart,
slice/.style={line cap=round, line join=round, very thick,draw=white},
pie title/.style={font={}},
slice type/.style 2 args={
    ##1/.style={fill=##2},
    values of ##1/.style={}
}
]

\pgfdeclarelayer{background}
\pgfdeclarelayer{foreground}
\pgfsetlayers{background,main,foreground}

\newcommand{\pie}[3][]{
    \begin{scope}[#1]
        \pgfmathsetmacro{\curA}{90}
        \pgfmathsetmacro{\r}{1}
        \def\c{(0,0)}
        \node[pie title] at (90:1.3) {#2};
        \foreach \v/\s in{#3}{
            \pgfmathsetmacro{\deltaA}{\v/100*360}
            \pgfmathsetmacro{\nextA}{\curA + \deltaA}
            \pgfmathsetmacro{\midA}{(\curA+\nextA)/2}

            \path[slice,\s] \c
            -- +(\curA:\r)
            arc (\curA:\nextA:\r)
            -- cycle;
            \pgfmathsetmacro{\d}{max((\deltaA * -(.5/50) + 1) , .5)}

            \begin{pgfonlayer}{foreground}
                \path \c -- node[pos=\d,pie values,values of \s]{$\v\%$} +(\midA:\r);
            \end{pgfonlayer}

            \global\let\curA\nextA
        }
    \end{scope}
}

\newcommand{\legend}[2][]{
    \begin{scope}[#1]
        \path
        \foreach \n/\s in {#2}
        {
            ++(1.4cm,0) node[\s,legend box] {} +(3pt,2pt) node[legend label] {\n}
        }
        ;
    \end{scope}
}

\usepackage{cite}
\usepackage{array}
\usepackage[hyphens]{url}
\usepackage{hyperref}
\usepackage[hyphenbreaks]{breakurl}
\usepackage{cleveref}
\begin{document}
\title{dewolf: Improving Decompilation \\ by leveraging User Surveys}

\author{\IEEEauthorblockN{
Steffen Enders\IEEEauthorrefmark{1},
Eva-Maria C.\ Behner\IEEEauthorrefmark{1},
Niklas Bergmann\IEEEauthorrefmark{1},
Mariia Rybalka\IEEEauthorrefmark{1},
Elmar Padilla\IEEEauthorrefmark{1},\\
Er Xue Hui\IEEEauthorrefmark{2},
Henry Low\IEEEauthorrefmark{2}, and
Nicholas Sim\IEEEauthorrefmark{2}}
\IEEEauthorblockA{\IEEEauthorrefmark{1}{Fraunhofer FKIE, Germany. \{firstname.lastname\}@fkie.fraunhofer.de}}
\IEEEauthorblockA{\IEEEauthorrefmark{2}{DSO National Laboratories, Singapore. \{exuehui,lzhenghe,sweishen\}@dso.org.sg}}
}

\maketitle

\begin{abstract}
    Analyzing third-party software such as malware or firmware is a crucial task for security analysts.
Although various approaches for automatic analysis exist and are the subject of ongoing research, analysts often have to resort to manual static analysis to get a deep understanding of a given binary sample.
Since the source code of encountered samples is rarely available, analysts regularly employ decompilers for easier and faster comprehension than analyzing a binary's disassembly.

In this paper, we introduce our decompilation approach \textit{dewolf}.
We developed a variety of improvements over the previous academic state-of-the-art decompiler and some novel algorithms to enhance readability and comprehension, focusing on manual analysis.
To evaluate our approach and to obtain a better insight into the analysts' needs, we conducted three user surveys.
The results indicate that dewolf is suitable for malware comprehension and that its output quality noticeably exceeds Ghidra and Hex-Rays in certain aspects.
Furthermore, our results imply that decompilers aiming at manual analysis should be highly configurable to respect individual user preferences.
Additionally, future decompilers should not necessarily follow the unwritten rule to stick to the code-structure dictated by the assembly in order to produce readable output.
In fact, the few cases where dewolf already cracks this rule lead to its results considerably exceeding other decompilers.
We publish a prototype implementation of dewolf and all survey results on GitHub~\cite{github2022dewolf,survey-results}.
\end{abstract}

\section{Introduction}
The number of malware-related incidents is steadily increasing, and with it is the workload of security analysts to analyze provided binary samples~\cite{avtest}.
Typically, most security analysts heavily rely on manual analysis to achieve their given objective, including a deep understanding of a binary's capabilities.
During this usually non-automated analysis, or more precisely manual static analysis, a \emph{decompiler} is a crucial tool to derive a source code-like representation of a given binary.
The comprehension of decompiled code is typically far less complicated than the binary's disassembly.
Therefore, a decompiler may significantly accelerate and ease analysis, rendering it an essential tool for binary analysis in general.
However, due to the information loss during compilation, a decompiler can not entirely revert the compilation process and has to make compromises that fit the developers' objective.
Consequently, most approaches for decompilation are developed specifically for a certain use-case or with a particular focus, such as aiming at recompilability for further automated processing~\cite{verbeek2020sound,brumley2013native}.
Although using a decompiler is an essential part of the static analysis workflow~\cite{votipka2020observational,yong2021inside}, only a few recent academic publications on decompilation focus on \emph{human} analysis~\cite{gussoni2020comb,yakdan2015no}.
Besides, the recent DARPA Cyber Grand Challenge showed that the performance of autonomous binary analysis is still far behind the abilities of reverse engineering experts~\cite{DARPA}.
Although researchers like Mantovani et al.~\cite{mantovani2022remind} try to understand how human analysts approach unknown binaries, their research is still in its very early stages.
We still deem manual analysis as one of the most important applications for decompilation.
As a consequence, we specifically aim at enhancing the output quality in terms of comprehensibility and readability to speed up manual analysis.

In this paper, we propose a novel approach for decompilation called \emph{dewolf}.
We base our approach on the previous research state-of-the-art \dream (see~\Cref{sec:selection}) and introduce a wide variety of improvements.
For example, we revise the restructuring of \dream by enabling the use of continue statements for control-flow interruptions and by refining the for-loop recovery.
Furthermore, dewolf incorporates various common algorithms such as subexpression elimination as well as the elimination of dead code, loops, and paths.
Moreover, we introduce a novel approach for out-of-SSA, without case distinctions, that potentially reduces the number of variables, and a custom logic engine.
Finally, we present various readability improvements such as utilizing compiler idiom handling, eliminating redundant casts, and improving the representation of constants.
Although some improvements include seemingly minor modifications, such as the display format of constants in the output, they lead to considerable enhancements of the output quality for human analysts according to our evaluation.

Besides, we achieve a remarkable level of stability for our \emph{research} decompiler using continuous reliability and correctness testing (see~\Cref{sec:testing}).
Overall, our developed decompiler is more stable than prototype implementations provided for other research decompilers such as \dream, while also considerably exceeding the code quality of previous approaches.
We publish the prototype implementation of dewolf as Binary Ninja plugin~\cite{github2022dewolf} to allow future research in the area of decompilation or related fields.

Apart from this technical contribution, we discuss the results of three user surveys we conducted between the years~2020 and~2021 (see~\Cref{sec:evaluation}).
Because we focus on a human-readable decompiler output, user surveys seem like the most reasonable choice for evaluation even given their natural limitations (see~\Cref{sec:eval:limitations}).
During those surveys, we studied various aspects including user preferences regarding decompiler output, weakness identification of current approaches, and the evaluation of dewolf regarding comprehension.
Each survey contributed to the identification of problems that we approached in~\Cref{sec:approach}, such as developing a custom logic engine in order to simplify logic expressions in the output.
Finally, in the third user survey, we included a comparison with other state-of-the-art decompilers to assess the significance of our approach.
Overall, we received~135 complete responses, with each survey having at least~37 participants and gathering many valuable results and insights.
Although our initial re-implementation of \dream was already very close to the state-of-the-art decompilers Ghidra and Hex-Rays, the survey results for dewolf show significant improvements.
More precisely, the final survey results indicate that dewolf's output for the provided function is clearly favored over others and ranked best by more than four out of five participants.
Furthermore, the survey results illustrate that dewolf's output is indeed suitable for malware comprehension.

One key finding of the user surveys suggests that decompilation should consider bolder approaches to reconstruct the control-flow.
Apparently, most current decompilers try to construct the control-flow similar to the structure dictated by the given assembly, e.g., only reconstructing a switch if a jump-table exists.
However, a less conservative reconstruction can enhance the human readability of the decompiled code tremendously, even if the results deviate from a structural translation of the disassembly.
Indeed, decompilation approaches like revng-c and \dream already apply certain reconstructions that do not necessarily originate from the considered assembly, such as copying instructions or introducing conditions to avoid goto-statements.
Nonetheless, our survey results suggest that decompilation approaches could go much further to improve readability leading to a lot of potential for future research while still retaining semantic equivalence.

To summarize, we make the following three contributions:
\begin{enumerate}
    \item A new decompilation approach, consisting of multiple individual readability improvements, whose output quality considerably exceeds \dream as well as Ghidra and Hex-Rays in certain aspects.
    \item Open-sourcing a prototype implementation of dewolf on GitHub~\cite{github2022dewolf} to allow future research and accessible development of new decompilation approaches.
    \item Publishing the entire user survey results to allow other researchers insight into our participants' perception of favorable code constructs and output.
\end{enumerate}
\section{Related Work}
\label{sec:related_work}

One of the earliest and most relevant publications in the field of decompilation is the PhD thesis of Cifuentes~\cite{cifuentes1994reverse}.
In this thesis, Cifuentes proposed a modular decompiler architecture where the actual decompilation algorithms, such as control- and data-flow analysis, are decoupled from those depending on the input binary architecture or the output high-level language.
To the best of our knowledge, most modern decompilers follow the proposed principle of a modular design.
The author also discussed general decompilation challenges and introduced algorithmic approaches to address various of them.
For example, Cifuentes proposed recovering high-level constructs via interval analysis, which allows mimicking the nesting order.
Here, a set of predefined patterns for known constructs is used to restructure an interval to a single node.

Emmerik~\cite{van2007static} suggested the usage of the static single assignment form (SSA-form) of a program to facilitate data-flow analysis needed for decompilation.
The author argued that classical data-flow algorithms, such as expression propagation, can be significantly simplified and sped up when applied to the program in SSA-form.
While transforming a program into its SSA-form is straightforward, finding an optimal algorithm for an \emph{out-of-SSA} transformation is not trivial~\cite{sreedhar1999translating}.
Nevertheless, many existing intermediate languages support both a SSA- and a non-SSA-form to facilitate data-flow analysis algorithms, including those introduced by Binary Ninja.

There are several academic decompilers that each focus on addressing specific decompilation challenges.
One central research field is the recovery of high-level control-flow constructs such as loops and if-else conditions, also known as control-flow restructuring or recovery.
The goal is to produce \emph{structured} code consisting of proper control-flow constructs and containing as few \emph{goto}-statements as possible to increase the comprehensibility of the decompiled code.

The Retargetable Decompiler (RetDec)~\cite{RETDEC_AVG07_15} also utilizes a modular design.
To disassemble the binary, K\v{r}oustek et al.\ use capstone~\cite{capstone}, and LLVM IR~\cite{llvm-ir-reference} as an intermediate language to minimize the overhead of adding support for new architectures.
Overall, there are many other publications regarding RetDec addressing different aspects such as compiler idiom handling~\cite{RETDEC_WAPL_13,RETDEC_KROUSTEK_PHD_15}.

Another novel research approach for decompilation is \mbox{RevEngE}~\cite{botacin2019revenge}.
Instead of decompiling the whole sample, they developed a \emph{debug-oriented decompilation} approach.
More precisely, during debugging, the analyst can decompile specific code regions.
Furthermore, it is possible to decompile regions more than once with different parameters.

Engel et al.~\cite{engel2011enhanced} improved structural analysis and extended it to handle C-specific control statements such as \emph{break}, \emph{continue} and \emph{return}.
They call this extension to structural analysis \emph{single entry single successor (SESS) analysis}.
One goal of this extension is to reduce the number of goto-statements because they argue that as few gotos as possible improve readability.
Additionally, to reduce the number of goto-statements, the authors define so-called \emph{tail regions}, which are regions that implicitly define an exit edge and its successor.
The authors also use a \emph{combined-condition pattern} which combines two if-statements having one common branch, resulting from conditions such as \mintinline{c}{if(A & B)}.

Schwartz et al.~\cite{brumley2013native} proposed \emph{iterative refinement} as an alternative for structural analysis with a focus on soundness.
They remove control-flow edges that impede existing algorithms by inserting a goto to allow the restructuring to continue.
Additionally, they introduced the \emph{semantic preservation} technique to guarantee that the original meaning of a program remains unchanged during restructuring.
They also suggested \emph{correctness} as a metric of decompiled output quality.
The approach was implemented in the Phoenix decompiler that is based on the BAP binary analysis framework.

Yakdan et al.~\cite{yakdan2015no} developed \emph{pattern-independent control-flow restructuring}, a novel algorithm to recover C control-flow constructs without generating any goto-statements.
In contrast to previous approaches, their algorithm does not rely on predefined patterns for control-flow constructs.
Instead, it utilizes conditions based on the reachability of nodes in the control-flow graph to infer the corresponding high-level constructs.
However, this also requires loops having a single entry and exit point.
Because not all real-world samples fulfill this requirement, the authors introduced a \emph{semantic-preserving transformation} technique that converts so-called \emph{abnormal loops} into their single-entry-exit equivalent.
Although such transformations allow subsequent goto-free restructuring, they lead to more complex control-flow graphs and ultimately longer output.
The authors implemented a decompiler prototype called \dream as an IDA Pro~6.5 plugin and further improved it by introducing readability enhancements.
They also conducted a user survey to evaluate their approach~\cite{yakdan2018human}.

The revng-c decompiler~\cite{gussoni2020comb} is implemented on top of the revng binary analysis framework and introduces another new yet partially similar approach to produce goto-free output.
The authors argue that the main obstacle to high-quality human-readable decompilation is that complex control-flow graphs contain a lot of tangled paths.
Consequently, they \emph{comb} complicated paths in the control-flow graph by duplicating particular graph nodes, which are either single basic blocks or already restructured and collapsed regions.
The resulting output is indeed goto-free and has a less tangled control-flow, but it may contain duplicated code snippets.
In order to avoid extensive copying of large regions, the authors perform \emph{untangling} which is the duplication of specific paths leading to the exit of the control-flow.
When untangling is necessary, it is performed before combing to duplicate fewer nodes.
Presumably, there is no universally valid guideline stating whether duplicated code should be preferred over a more complex control-flow as this may often depend on the given function and the analyst's personal preference.

In contrast to research approaches, commercial decompilers strongly focus on providing a stable environment and usability features to facilitate manual and automated binary analysis.
The Hex-Rays decompiler~\cite{hexrays} integrated into IDA Pro can be considered the de-facto industrial standard.
Although a few talks discuss some of Hex-Rays decompilation techniques~\cite{guilfanov2008decompilers,guilfanov2018decompilers}, the main decompilation approaches and algorithms remain unpublished.
On the other hand, Ghidra~\cite{ghidra}, the reverse engineering framework developed by the NSA, is widely considered to be a free alternative to Hex-Rays and the state-of-the-art open-source decompiler.
Instead of advancing the field of decompilation in general, Ghidra aims to ease decompilation for manual analysis.
While the developers did not scientifically publish any of their utilized approaches, both the documentation and source code are available on GitHub~\cite{github2022ghidra}.
Moreover, there are other less commonly mentioned decompilers besides Ghidra and Hex-Rays including FoxDec, JEB, angr, radare2's r2dec, Rizin's Cutter, and Binary Ninja's Pseudo C.

In the past few years, several decompilation approaches based on (recurrent) neural networks (NN) or neural machine translation have been proposed~\cite{katz2018using,katz2019towards,liang2021neutron}.
Although these approaches are allegedly language-independent, they are typically limited to short code snippets due to network design and memory requirement restrictions.
Most authors argue that writing and maintaining conventional or traditional decompilers is slow, costly, unscalable, or requires a high level of expert knowledge.
However, neural approaches often rely on sophisticated pre- and post-processing to add language domain knowledge to the model~\cite{katz2018using,katz2019towards} or even use traditional control- and data-flow recovery to improve the output~\cite{liang2021neutron}.

The most critical issue with approaches employing NNs or similar is that models typically learn the appearance of the code rather than capturing semantic equivalence.
Even though there is a loss of information during compilation which could justify not relying on semantics, the resulting binary still contains enough information to be translated in a semantic preserving way to the target language by exact and well-researched algorithms.
Additionally, finding a bug in incorrect decompilation is far more straightforward in a codebase employing a traditional approach than identifying the reasons for the model producing wrong results.
Finally, as current approaches illustrate, the utilization of NNs requires a lot of computational power to decompile even very simple code snippets.
However, decompilers are particularly essential for complex functions of which the assembly cannot be understood as easily as of simple functions.
Unfortunately, real-world programs and especially malware samples are typically rather large which renders current neural-network-based approaches unsuitable for reverse engineering of real-world malware.

\section{Approach}\label{sec:approach}

In this section, we introduce \emph{dewolf}, our improved approach for decompilation based on the previous state-of-the-art research approach \dream.
This section is structured as follows:
First, we will discuss our decompiler and intermediate language selection criteria in~\Cref{sec:selection}.
Next, in~\Cref{sec:method}, we will introduce our approach dewolf and describe its improvements over previous approaches.
Finally, we will conclude this section with additional details about some specifics of dewolf (see~\Cref{sec::logic,sec::out_of_ssa,sec:testing}).

\subsection{Framework Selection Criteria}\label{sec:selection}
Our goal is to improve the decompiler output in terms of readability by developing new and improving existing approaches.
The first step is to determine which decompilers or decompilation approaches are suitable for readability-focused research.
Previous studies have shown that gotos significantly aggravate analysts following the control-flow during static analysis~\cite{yakdan2016helping,gussoni2020comb,engel2011enhanced}.
Consequently, we decided to follow a goto-free approach and base our research on such an approach.
To the best of our knowledge, \dream and revng-c are the only two recent approaches producing goto-free output.
Therefore, these are the only two suitable decompilation approaches as a base for dewolf.

While both of those approaches use quite similar techniques for control-flow recovery and restructuring, \mbox{revng-c} uses code duplication in addition to structural variables to avoid goto-statements.
However, even though Gussoni et al.~\cite{gussoni2020comb} use untangling to reduce the amount of copied code, they still cannot prevent the duplication of large code blocks.
Although inserting a few lines of code may be beneficial when notably simplifying the control-flow, inserting large and potentially nested code snippets is often undesirable.
We ultimately opted to use the algorithms introduced by \dream over those from revng-c as code duplication may be unfavorable for the readability of the output.
Additionally, \dream achieved noteworthy results in their conducted user surveys while also focusing on human analysis.
Finally, starting with an approach that does not copy any instructions allows us to deliberately choose when, where, and how many instructions we copy in a future approach, potentially respecting a user's preference.

Because the openly published prototype implementation of \dream was not well-maintained and does not work on recent IDA versions, we decided to re-implement the algorithms introduced in the corresponding publications~\cite{yakdan2016helping,yakdan2018human}.
As part of the adoption, we want to base our approach on an intermediate language of an existing binary analysis framework due to the following reasons.
First, this avoids re-implementing well-known low-level algorithms such as translating a given function to SSA-from.
Second, this also ensures that our approach is platform-independent.
We decided not to integrate our approach into an existing decompiler-framework such as Ghidra or Hex-Rays because this usually impairs their workflow and requires extensive implementation efforts independent from the approach itself.

There are multiple intermediate languages and related lifters that can lift a given binary file each having their advantages and limitations~\cite{llvm-ir-reference,dullien2009reil,guilfanov2018decompilers}.
For our research, we decided on the following selection criteria:
(i) An accessible and comprehensive API to allow fast prototyping,
(ii) an included SSA-form for the intermediate language to assist data-flow analysis,
(iii) typed variables instead of registers as well as the elimination of stack-usages to further backup the platform-independence of our approach,
(iv) function call parameters linked to each function call,
and (v) a well-maintained framework.
Given these criteria, we opted to use the Medium Level Intermediate Language (MLIL) from Binary Ninja in SSA-form.
The MLIL allows us to develop decompilation approaches platform-independently while still supporting all platforms supported by Binary Ninja.
Additionally, MLIL implements our desired abstractions like variables instead of registers and offers a well-maintained and -documented API.
The impact of Binary Ninja on our results is limited to basic and well-known algorithms described in the selection criteria.

\subsection{Methodology}
\label{sec:method}
In the remainder of this section, we only describe differences or improvements we made over the existing approach.
All details not explicitly mentioned in this section are equal or at least similar to the publications related to \dream~\cite{yakdan2015no,yakdan2016helping}.
For instance, the control-flow restructuring has been kept identical except for some minor adjustments described below and some improvements of the switch-recovery.
As already mentioned, the prototype implementation of \dream as IDA Pro plugin is quite fragile and lacks support for~64~bit binaries.
Consequently, we re-implemented its algorithms as a Binary Ninja Plugin~\cite{binaryninja} utilizing the medium-level intermediate language (MLIL) in SSA-form as a starting point.

\pgfmathsetlengthmacro\arrowwidth{0.28\linewidth}
\pgfmathsetlengthmacro\arrowheight{1cm}

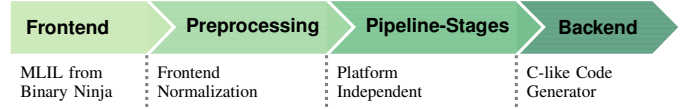
\begin{figure}[tb]
    \begin{adjustbox}{width=\linewidth}%
        \centering
      \trimbox{.2cm 1.5cm .1cm -.2cm}{
    \begin{tikzpicture}[
        >=latex,
        myshade/.style={
          shade, top color=#1!65,  bottom color=#1!65,
        },
      ]
        \matrix (m) [matrix of nodes, nodes in empty cells,
          nodes={draw=none, anchor=north west, minimum width=\arrowwidth,},
          column sep=0mm, row sep=0mm,
          row 1/.style={minimum height=\arrowheight,
            font=\hspace{5mm}\sffamily\bfseries, text=black
          },
          row 2/.style={anchor=north west,
            text width=\arrowwidth-0.4cm, font=\small,
            minimum height=\arrowheight, text depth=\arrowwidth-0.2cm,inner sep=2mm
        },
        ]{
          \hspace{-2.5em}Frontend & \hspace{.3em}Preprocessing & \hspace{.3em}Pipeline-Stages & Backend \\
          MLIL from Binary Ninja & Frontend \hspace{1em} Normalization & Platform \hspace{2em} Independent & C-like Code Generator \\
        };

          \begin{scope}[on background layer]
            \path[myshade=colorg5, draw=white, ultra thick, on background layer] (m-1-4.north west)
              -- (m-1-4.north east) -- +(0.2*\arrowwidth,-0.5*\arrowheight)
              -- (m-1-4.south east) -- (m-1-4.south west) --cycle;
            \path[myshade=colorg4, draw=white, ultra thick, on background layer] (m-1-3.north west)
            -- (m-1-3.north east) -- +(0.2*\arrowwidth,-0.5*\arrowheight)
            -- (m-1-3.south east) -- (m-1-3.south west) --cycle;
            \path[myshade=colorg3, draw=white, ultra thick, on background layer] (m-1-2.north west)
            -- (m-1-2.north east) -- +(0.2*\arrowwidth,-0.5*\arrowheight)
            -- (m-1-2.south east) -- (m-1-2.south west) --cycle;
            \path[myshade=colorg2, draw=white, ultra thick, on background layer] (m-1-1.north west)
            -- (m-1-1.north east) -- +(0.2*\arrowwidth,-0.5*\arrowheight)
            -- (m-1-1.south east) -- (m-1-1.south west) --cycle;
        \end{scope}
        \draw[gray, dotted, ultra thick] (m-1-3.south east) -- ($(m-1-3.south east) + (0,-1)$);
        \draw[gray, dotted, ultra thick] (m-1-2.south east) -- ($(m-1-2.south east) + (0,-1)$);
        \draw[gray, dotted, ultra thick] (m-1-1.south east) -- ($(m-1-1.south east) + (0,-1)$);
      \end{tikzpicture}}
    \end{adjustbox}
    \caption{The four phases of the dewolf decompiler.}
    \label{fig:approach:pipeline}
\end{figure}

\Cref{fig:approach:pipeline} shows the structure of dewolf.
Similarly to the decompiler structure proposed by Cifuentes~\cite{cifuentes1994reverse}, dewolf is structured into the four phases \emph{frontend}, \emph{preprocessing}, \emph{pipeline stages}, and \emph{backend}.
While frontend and backend are congruent with the design by Cifuentes, we utilize a preprocessing phase to normalize the frontend's output, more specifically, the output produced by Binary Ninja in our prototypical implementation.
The majority of our improvements to enhance the decompiler output or stability over \dream are implemented as pipeline stages and are platform-independent, like the universal decompiling machine proposed by Cifuentes.

Generally, we introduce three types of improvements:
First, we include commonly known or already published algorithms not yet implemented by \dream, e.g., using \texttt{continue} statements in loops.
Second, we improved some of the algorithms already integrated into \dream, such as stability improvements for for-loop recovery.
Finally, we introduce a few completely new algorithms, like a novel approach for out-of-SSA using graph coloring.
Two of the more complex improvements will be discussed more thoroughly in~\Cref{sec::logic,sec::out_of_ssa}.
To illustrate the improvements, \Cref{fig:comparison} shows an example function that has been decompiled with the initial re-implementation of \dream and also with the most recent version of dewolf.
The introduced improvements have a significant impact on the readability and the structure of the output, at least for this particular function.

In the remaining part of this section, we will give a short overview of achieved improvements over \dream and shortly summarize each improved algorithm.
Because most improvements are either well-documented elsewhere or relatively small, we will not discuss most of them in more detail.
Since we have published our prototype on GitHub~\cite{github2022dewolf}, anyone interested in more specific details of the introduced algorithms can look up the well-documented source code.

\begin{figure}[tb]
  \begin{subfigure}[l]{\linewidth}
    \begin{minted}[autogobble,bgcolor=my-bg,linenos,numbersep=4pt, breaklines,fontsize=\scriptsize]{c}
      void xor(char* s1, size_t slen, char* key, size_t keylen){
          printf("%lu\n", slen);
          for(int i=0; i<slen; ++i){
              s1[i] = s1[i] ^ key[i%keylen]; } }
    \end{minted}
    \subcaption{Source code of an example function.}
  \end{subfigure}
  \begin{subfigure}[l]{\linewidth}
    \begin{minted}[autogobble,bgcolor=my-bg,linenos,numbersep=4pt, breaklines,fontsize=\scriptsize]{c}
      int64_t xor(int64_t arg1, int64_t arg2, int64_t arg3, int64_t arg4){
          printf("%lu",arg2);
          int32_t var_c = 0x0L;
          while(true){
              int64_t rax_14 = var_c;
              if ((arg2 <= var_c)){ break; }
              char* rax_4 = (arg1 + var_c);
              uint64_t rsi_1 = *rax_4 ;
              int64_t rax_6 = var_c;
              int64_t rdx_1 = 0x0L;
              char* rax_9 = (arg3 + ((rdx_1:rax_6) % arg4));
              uint64_t rcx = *rax_9;
              int64_t rax_12 = (arg1 + var_c);
              uint64_t rdx_4 = (rsi_1.esi ^ rcx.ecx);
              *rax_12 = rdx_4.dl;
              int32_t var_c = (var_c + 0x1L); }
          return rax_14; }
    \end{minted}
    \subcaption{dewolf's output of initial re-implementation of \dream.}
  \end{subfigure}
  \begin{subfigure}[l]{\linewidth}
    \begin{minted}[autogobble,bgcolor=my-bg,linenos,numbersep=4pt, breaklines,fontsize=\scriptsize]{c}
      long xor(void * arg1, long arg2, long arg3, long arg4){
          int i;
          int var_1;
          void * var_0;
          printf(/* format */ "%lu\n", arg2);
          for (i = 0; arg2 > i; i++){
              var_0 = arg1 + i;
              *var_0 = *var_0 ^ *(arg3 + (i + (0L << 0x40)) % arg4); }
          var_1 = i;
          return var_1; }
    \end{minted}
  \subcaption{dewolf's output which is more concise and readable.}
  \end{subfigure}
  \caption{Comparison between the output of our \dream re-implementation and dewolf.}
  \label{fig:comparison}
\end{figure}

\subsubsection{Liveness Analysis}
We use liveness analysis to construct the interference graph that we need in various algorithms of dewolf.
For example, we use it to create the dependency graph used during the out-of-SSA transformation.
In our approach, we utilize liveness sets by exploring paths from variable usages according to Brandner et al.\ (Algorithm~4)~\cite{brandner2011computing}.

\subsubsection{Switch Variable Detection}
To enhance the comprehension of complicated control-flows, programmers frequently use switch statements.
However, in assembly generated by modern compilers, switch control-flows are often implemented using indirect jumps and jump-tables.
To recover switch control-flows in those cases, we reconstruct the original switch variable by tracing the variable used to calculate the jump-offset until we reach its definition.
Eventually, we can use this information to restructure a region using a switch to increase the readability.

\subsubsection{Subexpression Elimination}
After propagating expressions, there still may be certain expressions that occur multiple times.
Unfortunately, depending on the complexity of such expressions, comprehending them multiple times is quite time-consuming for the analyst.
During subexpression elimination, we identify expressions that are already defined and used in other instructions as well as expressions that occur multiple times.
Using the dominator tree, we find the closest possible position to its usages for insertion to avoid unnecessary interference between variables.
Eventually, we insert a definition for the common subexpression at the identified location and replace all its occurrences.

\subsubsection{Constant Representation}
In general, constant values are often displayed as hex-numbers in the decompiler output.
However, as constant values can represent integers, characters, addresses, or else, a hexadecimal number might not always be the most suitable representation.
Consequently, we approximate the most likely representation depending on the constant's type and value to enhance the comprehension of the decompiler output.

\subsubsection{For-loop Recovery}
While \dream already included transformation rules from while-loops into for-loops, the introduced requirements to transform a given loop are actually quite strict.
Because these strict rules result in only very few for-loop transformations in real-world samples, we developed a more stable for-loop recovery.
Therefore, we can recover for-loops in many additional cases by altering and relaxing the rules proposed by \dream.
For example, in contrast to \dream, we allow that the incrementation of the loop counter can be another than the loop body's \emph{last} instruction as long as the defined and used variables remain unchanged before the next loop-iteration.

\subsubsection{Compiler Idioms}
Unresolved compiler idioms can be very challenging for the comprehension of a given function as they typically translate to complicated arithmetic calculations rather than the intended high-level expression.
While \dream did not consider compiler idioms, we decided to include algorithms to revert them.
More specifically, we opted to integrate a very recently published approach for compiler idiom handling~\cite{enders2021pidarci} which uses automatically generated assembly patterns to annotate and revert the most common idioms.

\subsubsection{Array Access Detection}
In assembly language or intermediate languages, the concept of arrays does not exist.
Instead, an array element access is represented as a dereference operation to a variable plus an offset, i.e., \mintinline{c}{*(base + valid_offset)}.
We find such candidates and mark them as potential array element access.
Furthermore, we compute the base and the index, depending on the offset and the array size.
Eventually, we discard candidates not fulfilling certain requirements, such as a consistent array size.
The decompiler output displays dereference operations marked as array access as \mintinline{c}{base[i]}.

\subsubsection{Elimination of Dead Paths and Loops}
After propagating expressions, some branch conditions may be rendered unfulfillable (or the opposite) and therefore introduce impossible paths ultimately leading to unreachable code.
Certainly, analyzing dead code usually is not part of the analyst's objective and may significantly decrease efficiency.
We identify such conditions and then eliminate \emph{dead} paths or paths that are unreachable.
Similarly, we can also remove dead loops which occur when the loop condition is always unsatisfied during the first loop entry.
Overall, both algorithms remove unreachable basic blocks and can significantly reduce analysis workload.

\subsubsection{Elimination of Redundant Casts}
Binary programs produced by modern compilers are usually heavily optimized to make efficient use of the available registers.
Since many assembly languages allow to access only parts of a given register, various cast operations can be introduced when lifting the assembly code to a more high-level intermediate representation.
After propagating expressions, we may end up with many nested casts, some being semantically irrelevant.
To declutter the output and considerably increase its readability, we introduce a set of rules for when a cast operation can be omitted due to irrelevance.

\subsubsection{ Continue in loops}
The \dream approach solely uses break-statements to indicate when the control-flow may exit a given loop.
In contrast, we improved the control-flow restructuring by also using continue-statements to decrease the nesting depth where possible.
More specifically, during the pattern-independent restructuring proposed by \dream, we do not exclusively add break-nodes on edges used to exit the loop.
Instead, we additionally add continue-nodes on every edge having the loop-head as a sink.

\subsection{Custom Logic Engine} \label{sec::logic}

Logic expressions in binary executables are often strongly optimized to the target architecture and therefore are not always an intuitive translation of the original expressions.
As a consequence, decompilers often employ a logic engine to simplify such expressions~\cite{yakdan2018human} but also to, e.g., identify unreachable code during the elimination of dead loops/paths or during control-flow restructuring.
However, the amount of logic engines featuring full support for logic on bit-vectors is quite limited.
While the \dream approach uses SymPy and z3 for condition simplification, it turned out that SymPy is too slow even when simplifying moderately long conditions.
The z3 Theorem Prover is an open-source logic engine supporting satisfiability modulo theories (SMT) and bit-vector logic.
Considering the vast number of forks on github~\cite{githubz3} and citations of their initial publication~\cite{de2008z3}, z3 can be considered the most relevant theorem prover and state-of-the-art.

Unfortunately, although being the state-of-the-art SMT solver, z3 is not optimal for our purpose either.
Although being a full-fledged theorem prover and offering full support for bit-vectors, some design choices of z3 are unsuited to model popular processor architectures such as i386.
For example, z3 favors signed operations over unsigned operations.
While this seems like a reasonable choice, it may lead to unexpected simplification results when used for decompilation.
For example, version~4.8.10 may simplify an unsigned operation into a conjunction of two operations, e.g., with all operands represented by~32-bit vectors:
\[
  a {\leq}_u 25 \models a[5:31] = 0 \land a[0:4] \leq 25.
\]
Even though this simplification does not alter the logic, it raises the complexity and length of the output considerably.
This lack of simplification is particularly relevant since unsigned operations are predominant when handling pointers on any architecture.
Additionally, z3 has problems simplifying conditions that often occur during the restructuring, e.g., z3 is not able to make the following simplification:
\[
  (a = 1 \lor a = 2 \lor a = 3 \lor a = 4) \models 1 \leq a \leq 4.
\]

Without a doubt, the complexity of logic statements as well as branching conditions has a direct impact on the readability of the decompiled code.
Because we noticed several simplification problems, and had significant obstacles with signed operations, we eventually opted to develop a graph-based logic engine fitting our requirements.
Although it only provides a small subset of the functionality provided by z3, we designed it for the particular use case of modeling logic expressions generated from assembly language.
More specifically, using a custom logic engine allows us to include exactly the simplifications we need during decompilation.

The custom engine utilizes graphs to model logic formulas and defines simplification methods for various common combinations of different operations.
We store each formula as part of a graph representing all conditions for a given context which can contain multiple terms and clauses.
Further, each operation and bit vector (representing a variable or constant) is represented by a graph node.
Finally, we represent the relation between the operations and bitvectors via directed-edges, either definition edges linking a variable to its definition or operand edges connecting an operation with its operands.

Using our graph representation, we can define a lightweight form of equality based on graph coloring algorithms, also used for graph similarity~\cite{mutzelSTACS2019}.
Besides, it offers many practical advantages, e.g., we do not need sophisticated methods to deal with the ordering of operands or to copy and replace terms in an expression.
Finally, we can traverse edges to compute the graph predecessors or successors to view relations easily, e.g., between subexpressions.
We open-sourced the logic engine alongside the decompiler on GitHub~\cite{github2022dewolflogic}.

\subsection{Improved Out-of-SSA} \label{sec::out_of_ssa}

Many data-flow-algorithms are applied to a program in SSA-form because it simplifies many algorithmic challenges~\cite{van2007static}.
However, because in SSA-form each variable is defined exactly once, basic blocks with multiple entries may cause variables to have a different value depending on the control-flow.
To resolve this problem, \emph{$\varphi$-functions} are used to assign variable values depending on the control-flow.
Conceptually, all $\varphi$-functions are executed simultaneously in each basic block.
However, there is no analogous concept of $\varphi$-functions in the C-language.
Thus, we have to eliminate them at some point to obtain valid C output.
Unfortunately, removing $\varphi$-functions is not trivial due to the lost-copy and swap-problem~\cite{sreedhar1999translating}.
Hence, it is no surprise that there is still ongoing research about this topic~\cite{brandner2011computing, sreedhar1999translating}.
However, most of these algorithms are rather complicated and distinguish many different cases.
Therefore, we developed a more straightforward algorithm.
One benefit of our approach is that we do not distinguish between multiple cases depending on the usage of variables in the $\varphi$-functions of a basic block.

\begin{figure*}[tb]
    \tikzset{myptr/.style={decoration={markings,mark=at position 1 with %
{\arrow[scale=0.75,>=triangle 60]{>}}},postaction={decorate}}}
    \begin{subfigure}[b]{0.33\textwidth}
        \centering
        \begin{tikzpicture}[transform shape, scale=0.8]
            \node[draw, rectangle, text width=2cm] (P1) at (0, 0) {%
                \texttt{%
                \varz{3} = bar()\\
                }
            };
            \node[draw, rectangle, text width=2.7cm] (P2) [below =0.5cm of P1] {%
                \texttt{%
                \vary{2} = $\varphi$(\vary{1},\varz{2}) \\
                \varz{2} = $\varphi$(\varz{1},\vary{2}) \\
                \varu{2} = $\varphi$(0x1,\varu{1}) \\
                if(\varu{2} $\leq$ 14)\\
                }
            };
            \node[draw, rectangle, minimum height = 1.5em, text width=1.8cm] (P4) [below left = 0.5cm and -0.5cm of P2] {
                \texttt{return \vary{1}}
            };
            \node[draw, rectangle, minimum height = 1.5em, text width=2.3cm] (P3) [below right = 0.5cm and -0.5cm of P2] {
                \texttt{%
                    \varu{1} = \varu{2} + 0x1 \\
                }
            };
            \draw[myptr, very thick, draw=black!70] (P1) -- (P2);
            \draw[myptr, very thick, draw=Green] (P2.east) -| (P3.north);
            \draw[myptr, very thick, draw=Red] (P2.west) -| (P4.north);
            \draw[myptr, very thick, draw=black!70] (P3.west) -| (P2.south);
        \end{tikzpicture}
        \caption{Initial cfg}
    \end{subfigure}
    \begin{subfigure}[b]{0.33\textwidth}
        \centering
        \begin{tikzpicture}[transform shape, >=triangle 60]
            \begin{scope}
            \node[draw, circle] (y2) at (0, 0) {\vary{2}};
            \node[draw, circle] (z2) [left =of y2] {\varz{2}};
            \node[draw, circle] (u2) [right =of y2] {\varu{2}};

            \draw[draw=black!70,->, very thick] (y2) to [bend left] (z2);
            \draw[draw=black!70,->, very thick] (z2) to [bend left] (y2);
            \end{scope}

            \begin{scope}[yshift=-1cm]
            \node[draw, circle, dashed] (y2) at (0, 0) {\vary{2}};
            \node[draw, circle] (z2) [left =of y2] {\varz{2}};
            \node[draw, circle] (u2) [right =of y2] {\varu{2}};

            \draw[draw=black!70,->, dashed] (y2) to [bend left] (z2);
            \draw[draw=black!70,->, dashed] (z2) to [bend left] (y2);
            \end{scope}

            \begin{scope}[yshift=-2cm]
            \node[draw, circle] (yp2) at (0, 0) {\varyp{2}};
            \node[draw, circle] (z2) [left =of yp2] {\varz{2}};
            \node[draw, circle] (u2) [right =of yp2] {\varu{2}};

            \draw[draw=black!70,->, very thick] (yp2) -- (z2);
            \end{scope}
        \end{tikzpicture}
        \caption{$\varphi$ dependency graph, $y_2$ directed fvs.}
        \label{fig:ssa:dependency}
    \end{subfigure}
    \begin{subfigure}[b]{0.33\textwidth}
        \centering
        \begin{tikzpicture}[transform shape, scale=0.8, >=triangle 60]
            \node[draw, rectangle, text width=2cm] (P1) at (0, 0) {%
                \texttt{%
                \varz{3} = bar()\\
                }
            };
            \node[draw, rectangle, text width=2.7cm] (P2) [below =0.5cm of P1] {%
                \texttt{%
                \varyp{2} = $\varphi$(\vary{1},\varz{2}) \\
                \varz{2} = $\varphi$(\varz{1},\vary{2}) \\
                \varu{2} = $\varphi$(0x1,\varu{1}) \\
                \vary{2} = \varyp{2} \\
                if(\varu{2} $\leq$ 14)\\
                }
            };
            \node[draw, rectangle, minimum height = 1.5em, text width=1.8cm] (P4) [below left = 0.5cm and -0.5cm of P2] {
                \texttt{return \vary{2}}
            };
            \node[draw, rectangle, minimum height = 1.5em, text width=2.3cm] (P3) [below right = 0.5cm and -0.8cm of P2] {
                \texttt{%
                    \varu{1} = \varu{2} + 0x1 \\
                }
            };
            \draw[myptr, very thick, draw=black!70] (P1) -- (P2);
            \draw[myptr, very thick, draw=Green] (P2.east) -| (P3.north);
            \draw[myptr, very thick, draw=Red] (P2.west) -| (P4.north);
            \draw[myptr, very thick, draw=black!70] (P3.west) -| (P2.south);
        \end{tikzpicture}
        \caption{Removed circular dependencies.}
        \label{fig:ssa:removed}
    \end{subfigure}
    \begin{subfigure}[b]{0.33\textwidth}
        \centering
        \begin{tikzpicture}[transform shape, scale=0.8, >=triangle 60]
            \node[draw, rectangle, text width=3.6cm] (P1) at (0, 0) {%
                \texttt{%
                \varz{3} = bar(); \varyp{2} = \vary{1}; \\
                \varz{2} = \varz{1}; \varu{2} = 0x1 \\
                }
            };
            \node[draw, rectangle, text width=2.7cm] (P2) [below =0.5cm of P1] {%
                \texttt{%
                \vary{2} = \varyp{2}\\
                if(\varu{2} $\leq$ 14)\\
                }
            };
            \node[draw, rectangle, minimum height = 1.5em, text width=1.8cm] (P4) [below left = 0.5cm and -0.5cm of P2] {
                \texttt{return \vary{2}}
            };
            \node[draw, rectangle, minimum height = 1.5em, text width=2.3cm] (P3) [below right = 0.5cm and -0.8cm of P2] {
                \texttt{%
                    \varu{1} = \varu{2} + 0x1 \\
                    \varyp{2} = \varz{2} \\
                    \varz{2} = \vary{2} \\
                    \varu{2} = \varu{1} \\
                }
            };
            \draw[myptr, very thick, draw=black!70] (P1) -- (P2);
            \draw[myptr, very thick, draw=Green] (P2.east) -| (P3.north);
            \draw[myptr, very thick, draw=Red] (P2.west) -| (P4.north);
            \draw[myptr, very thick, draw=black!70] (P3.west) -| (P2.south);
        \end{tikzpicture}
        \caption{Lifted $\varphi$-functions}
        \label{fig:ssa:lifted}
    \end{subfigure}
    \begin{subfigure}[b]{0.33\textwidth}
        \centering
        \begin{tikzpicture}[transform shape]
            \node[draw, circle, fill=darkcerulean!20!white] (c_1) {\vary{2}};
            \node[draw, circle, fill=shamrockgreen!20!white] (c_2) [below right = 1 and 0.5 of c_1] {\varz{2}};
            \node[draw, circle, fill=brass!20!white] (c_3) [left =of c_2] {\varu{2}};
            \node[draw, circle, fill=antiquefuchsia!20!white] (c_4) [left =of c_1] {\varyp{2}};
            \node[draw, circle, fill=antiquefuchsia!20!white] (c_5) [right =of c_1] {\vary{1}};
            \node[draw, circle, fill=shamrockgreen!20!white] (c_6) [right =of c_2] {\varz{1}};
            \node[draw, circle, fill=brass!20!white] (c_7) [left =of c_3] {\varu{1}};

            \foreach \i/\j in {1/2,1/3,1/4,1/7,2/3, 2/4, 2/5, 3/4, 4/6}{
                \draw[thick] (c_\i) -- (c_\j);
            }
            \draw[-, thick] (c_6) to [bend left=20] (c_7);
        \end{tikzpicture}
        \caption{Colored interference graph.}
        \label{fig:ssa:colored}
    \end{subfigure}
    \begin{subfigure}[b]{0.33\textwidth}
        \centering
        \begin{tikzpicture}[transform shape, scale=0.8, >=triangle 60]
            \node[draw, rectangle, text width=2cm] (P1) at (0, 0) {%
                \texttt{%
                \varz{} = bar()\\
                \varu{} = 0x1 \\
                }
            };
            \node[draw, rectangle, text width=2.7cm] (P2) [below =0.5cm of P1] {%
                \texttt{%
                \vary{} = \varyp{}\\
                if(\varu{} $\leq$ 14)\\
                }
            };
            \node[draw, rectangle, minimum height = 1.5em, text width=1.8cm] (P4) [below left = 0.5cm and -0.5cm of P2] {
                \texttt{return \vary{}}
            };
            \node[draw, rectangle, minimum height = 1.5em, text width=2.3cm] (P3) [below right = 0.5cm and -0.8cm of P2] {
                \texttt{%
                    \varu{} = \varu{} + 0x1 \\
                    \varyp{} = \varz{} \\
                    \varz{} = \vary{} \\
                }
            };
            \draw[myptr, very thick, draw=black!70] (P1) -- (P2);
            \draw[myptr, very thick, draw=Green] (P2.east) -| (P3.north);
            \draw[myptr, very thick, draw=Red] (P2.west) -| (P4.north);
            \draw[myptr, very thick, draw=black!70] (P3.west) -| (P2.south);
        \end{tikzpicture}
        \caption{Resulting cfg.}
    \end{subfigure}
\caption{Illustration of each step from our out-of-SSA algorithm applied to an example cfg.}
\label{fig:ssa}
\end{figure*}
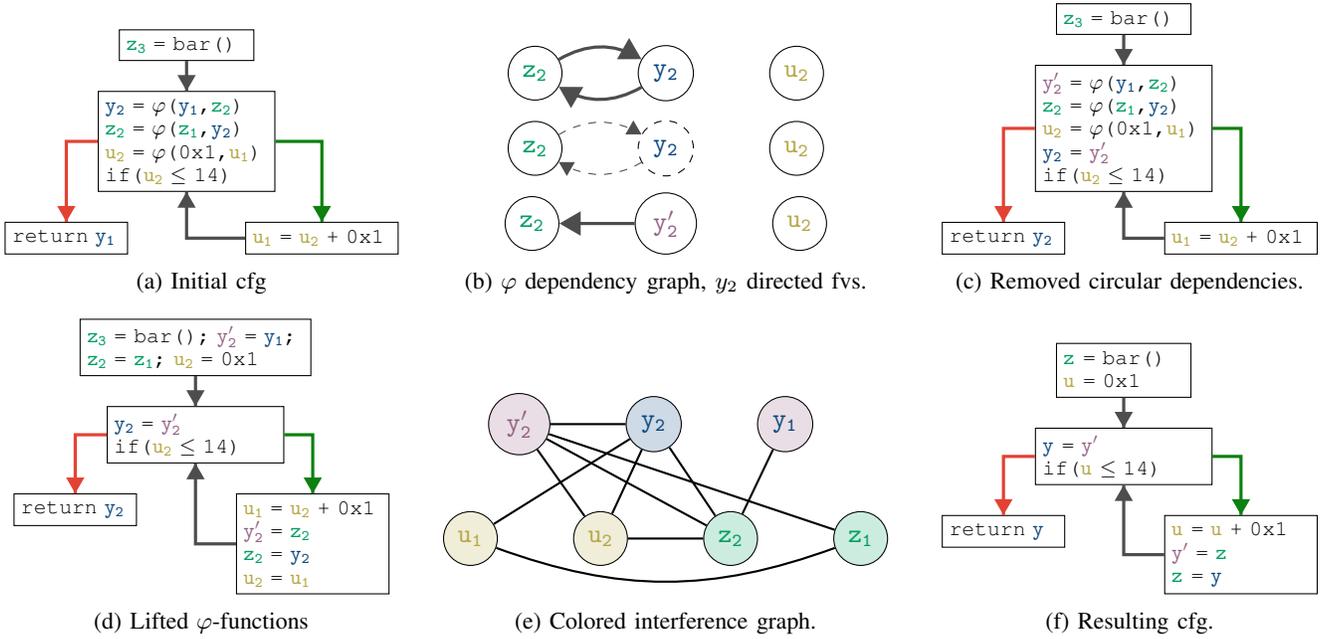

Our approach consists of three algorithms, namely \emph{removing circular dependencies} of $\varphi$-functions, \emph{lifting} $\varphi$-functions, and \emph{renaming} variables.
It is possible to remove the circular dependency and lift the $\varphi$-functions first, and then rename the variables or the other way around.
Similar to other approaches, we remove the circular dependency on $\varphi$-functions by inserting copies.
We argue that the $\varphi$-functions $\varphi_1, \varphi_2, \ldots, \varphi_l$ depend circularly on each other if they are all contained in the same basic block and if the variable, defined via $\varphi_i$ (for $1 \leq i \leq l$), is used in $\varphi_{i+1}$ (with $\varphi_{l+1}=\varphi_1$).
Hence, it is impossible to order $\varphi$-functions such that we execute them successively when they depend circularly on each other.
After resolving the circular dependencies, we can order the $\varphi$-functions into a possible execution order.
Given such an order, it is possible to move the instructions represented by the $\varphi$-functions to the predecessor blocks.
In the following, we will describe each of the three steps in more detail.
\Cref{fig:ssa} illustrates an overview of the approach applied to a cfg.

\subsubsection{Remove circular dependency}
As mentioned before, the goal of this algorithm is to insert copies such that we can execute the $\varphi$-functions one after another.
To find the variables for which we have to insert copies, we construct a dependency graph as shown in~\Cref{fig:ssa:dependency}.
This graph contains one vertex for each variable that is defined via a $\varphi$-function and we add an edge $(u, v)$ between two variables $u, v$ if $v$ is used in the $\varphi$-function defining $u$.
This means, having the $\varphi$-functions $u =\varphi(0, v)$ and $v = \varphi(y, z)$, we add the edge $(u, v)$ to the dependency graph.
The edge $(u, v)$ indicates that we have to execute the $\varphi$-function defining $u$ before the $\varphi$-function defining $v$ because we need the value of $v$ before the $\varphi$-functions reassigns it for the computation of $u$.
Now, a cycle in the dependency graph is equivalent to having $\varphi$-functions that depend circularly on each other.

To resolve the circular dependency between the $\varphi$-functions, we compute a directed feedback vertex set (directed fvs) in the dependency graph; a directed fvs is a set $X$ of vertices whose removal results in an acyclic graph.
Although computing a directed fvs is NP-hard~\cite{karp1972reducibility}, we can find such a set in constant time because we only have a constant number of vertices in each basic block and because we do this computation for each basic block independently.
Nevertheless, we only use an approximation to compute the directed fvs.
Now, for every $\varphi$-function $x = \varphi(\ldots)$ defining a variable $x$ contained in the directed fvs $X$, we replace the $\varphi$-function by $\mathrm{copy}\_x = \varphi(\ldots)$.
Furthermore, we add the instruction $x=\mathrm{copy}\_x$ after the last $\varphi$-function because variable $x$ must have the value of $\mathrm{copy}\_x$ after all as shown in~\Cref{fig:ssa:removed}.
Afterward, the dependency graph is acyclic because it does not contain any vertex of the set $X$, and all copies are isolated in the dependency graph.
Finally, we arrange the $\varphi$-function according to a topological order of the dependency graph, implying that we can execute them in this order.

\subsubsection{Lift $\varphi$-functions}
To apply this algorithm the circular dependency of the $\varphi$-functions must be resolved and the $\varphi$-functions must be ordered in a valid execution order.
Thus, this algorithm always runs directly after resolving the circular dependencies and sorts the $\varphi$-functions accordingly.

Since the $\varphi$-functions can be executed successively, we can "lift" the instructions represented by each $\varphi$-function to the predecessor basic blocks.
More precisely, for each $\varphi$-function $x_0 = \varphi(x_1, x_2, \ldots, x_l)$ we have a mapping that maps from each predecessor basic block to the variable or constant assigned to $x_0$ when the flow enters over this basic block.
We add the assignment $x_0 = x_i$ at the end of the basic block $\mathrm{BB}_i$ corresponding to $x_i$ when the basic block does not end with a branch.
Otherwise, we add a new basic block between the current basic block and $\mathrm{BB}_i$ and add the assignment in the new basic block.
After this step, all $\varphi$-functions are removed by moving the definitions to the predecessor basic blocks as shown in~\Cref{fig:ssa:lifted}.
The lifted assignments are ordered according to the order of the $\varphi$-functions.
After this step, the program is no longer in SSA-form.

\subsubsection{Rename Variables}
There are many possible ways to rename the variables.
The easiest renaming strategy is to assign each variable its name in SSA-form, i.e., the variable $v$ with label 4 gets name $v\_4$.
However, this leads to many variables and many copy assignments.

In general, when renaming variables, we only have to make sure that two variables solely get the same name when they do not interfere.
One possible way to achieve this is to use graph coloring.
More precisely, given any valid coloring of the interference graph like in~\Cref{fig:ssa:colored}, we can assign all variables of the same color class the same name.
Sadly, finding a vertex-coloring with a minimum number of colors is NP-hard~\cite{karp1972reducibility}.
However, there are many special cases where this problem is solvable in polynomial time, e.g., in chordal graphs, and there exist multiple heuristic and greedy algorithms.
Currently, we use lexicographical BFS to find the order in which we color the vertices.
We always use an existing color, if possible, and introduce a new color otherwise.
To get rid of as many copy-assignments as possible, we have some additional rules on how to choose between multiple possible existing colors.

As mentioned before, it is possible to apply the algorithms in two different orders.
When renaming variables first, the program is still in SSA-form which implies that the interference graph is chordal~\cite{hack2005interference}.
Thus, we can compute an optimal coloring in linear time using our coloring algorithm.
Although this initially leads to a minimal number of variables, starting with the renaming can produce more circular dependencies between the $\varphi$-functions.
Hence, we may have to insert more copies when resolving the circular dependencies.

On the other hand, the interference graph may no longer be chordal once we removed the $\varphi$-functions by removing the circular dependencies and lifting them.
Consequently, our algorithm may not compute an optimal coloring.
However, due to our additional rules on how to choose between the existing colors and the fact that the graph is nearly chordal, the result is still sufficient.

\subsection{Reliability and Correctness Testing}\label{sec:testing}

While developing the above improvements, we made significant changes to the decompilation process, each typically based on a small number of test functions.
However, it is impossible to consider the entire universe of binary functions when developing any new algorithm, resulting in bugs and crashes when decompiling other functions.
This section outlines our approach for decompiler testing and the strategies we used to improve dewolf's reliability and correctness.

\subsubsection{Decompiling a fixed set of functions}
In addition to handwritten test cases, we used Coreutils~8.32 and a subset of the LLVM test suite to find regressions and fix decompilation bugs.
We integrated tests for the Coreutils dataset to our continuous integration pipeline to proactively  identify problematic functions for new analyses.

\subsubsection{Fuzzing with Csmith}
We fuzzed dewolf with Csmith-generated functions~\cite{yang-csmith}, to find bugs that did not occur in the above set.
While this was most useful during early development, it did not help to identify subtler semantic issues.

\subsubsection{Differential fuzzing of re-compiled code}
We used differential fuzzing to test the semantic correctness of our lifter and transformations.
We limited our attempts for re-compilation of decompiled code to the simplest possible functions from the LLVM test suite and generated by Csmith, which
(i) do not call other functions,
(ii) do not use global variables, and
(iii) do not use structures or unions.
By excluding vast swathes of possible programs, we can focus on semantic transformations applied by dewolf.

We tested the originally compiled binaries against their decompiled and re-compiled counterparts.
To perform the tests, we generated harnesses for use with the Nezha framework~\cite{nezha2017}.
The differential fuzzing tests helped us to fix multiple integer promotion casting bugs as well as type recovery bugs which we would not have identified otherwise.
\Cref{fig:diff-fuzz} shows an example for an identified bug.

\begin{figure}
  \begin{subfigure}[l]{\linewidth}
    \begin{minted}[autogobble,bgcolor=my-bg,linenos,numbersep=4pt,breaklines,breakanywhere,fontsize=\scriptsize,highlightlines={3},highlightcolor=my-bg!90!black]{c}
      int test_case(int argc) {
        short s1 = (argc >= 3) ? argc : -769;
        unsigned short us2 = (unsigned short) s1;
        return us2; }
    \end{minted}
    \subcaption{
      Original source-code, adapted from the LLVM test suite~\cite{llvm-test-suite}, resulting in
      \mintinline{c}{test_case(69794316) == 64012} (2 bytes).
    }
  \end{subfigure}
  \begin{subfigure}[l]{\linewidth}
    \begin{minted}[autogobble,bgcolor=my-bg,linenos,numbersep=4pt,breaklines,breakanywhere,fontsize=\scriptsize,highlightlines={7},highlightcolor=my-bg!90!black]{c}
      unsigned long test_case(int arg1) {
        short var_0;
        if (arg1 > 2) var_0 = (short)arg1;
        else var_0 = -769;
        return (unsigned int)var_0; } // should be ushort
    \end{minted}
    \subcaption{
      Recompiled output from dewolf decompiler, resulting in
      \mintinline{c}{test_case(69794316) == 4294965772} (4 bytes).
    }
  \end{subfigure}
  \caption{
    \label{fig:diff-fuzz}
    A casting bug found using differential fuzzing.
    Both values are equal to -1524 in their respective signed variants, but leads to confusion when used as-is by a solver.
  }
\end{figure}

Our testing experience showed that fuzzing and differential testing are readily applicable to decompilers.
While we were only able to re-compile and differentially fuzz a small subset of randomly-generated functions, it has increased our confidence in the accuracy and correctness of dewolf.

\section{Evaluation}
\label{sec:evaluation}

As already mentioned, our improvements for decompilation approaches from the previous sections are heavily focused on enhancing decompilation for manual static analysis by humans.
In general, the quality of code can either be evaluated using quantitative measurements, i.e., code metrics, or by conducting user surveys.
Some of the most common code metrics are \emph{Source Line of Code}, \emph{Halstead Software Science}, \emph{ABC}, \emph{McCabe Cyclomatic Complexity}, \emph{Henry and Kafura's Information Flow}, and \emph{Nesting depth}.
However, a major downside of using such metrics for evaluation is that each metric favors a particular behavior which may not necessarily lead to a better code readability.
For example, the revng-c approach~\cite{gussoni2020comb} uses the McCabe Cyclomatic Complexity for their evaluation.
However, this metric favors their way of duplication without measuring its disadvantages.
When copying a node with $\geq 2$ predecessors and one successor such that each duplication has one predecessor, the code size may considerably increase whereas the cyclomatic complexity does not change.
When duplicating arbitrarily big basic blocks without any successor, the cyclomatic complexity may even decrease, although such duplicating obviously harms the code's readability and decreases its quality.

Even combinations of code metrics such as McCabe Cyclomatic Complexity \emph{and} Lines of Code are not sufficient to assess the readability since reducing the lines of code does not always help a human analyst.
For example, tweaking expression propagation to generate rather long lines of code can be quite difficult to understand for a human analyst although reducing the lines of code.
To conclude, each code quality metric considers only one aspect of the code and hence has advantages and disadvantages~\cite{bhatti2011automatic}.
Consequently, the most natural way to evaluate our approach in terms of readability is to conduct user surveys.
Principally, a user survey is \emph{the only} way to measure improvements in readability or comprehension for a human analyst, as they can be measured by comparing outputs or extracting a baseline from the answers.
Therefore, using an extensive survey to study analysts' preferences regarding code constructs or output formats is most appropriate.

As described in \Cref{sec:approach}, our decompilation approach consists of various improvements which we can either evaluate individually or collectively as a whole.
Although the individual and incremental evaluation would allow assessing the individual impact of each improvement, this would result in an unmanageable number of combinations which goes far beyond what we are able to evaluate with user surveys.
Additionally, we presume that the impact of each individual contribution is rather small and may not result in measurable improvements.
Consequently, we opted for an evaluation of our decompiler approach combining all introduced improvements.

During our research, we conducted a total of three user surveys, each with a focus on the evaluation of one of the following three aspects:
First, identifying the limitations of \dream, finding constructs preferrable by analysts, as well as relevant research questions.
Second, we evaluated whether the improved output of our approach is sufficient to comprehend a realistic malware sample since this is a crucial task of a decompiler.
Finally, we compared our approach to the open-source and commercial state-of-the-art decompilers Ghidra and Hex-Rays.
In general, each conducted user survey consisted of multiple parts and was of considerable length (see~\Cref{table:survey:overview}).
In addition to the already mentioned parts, we also included questions for self-assessment and feedback regarding the survey itself.
Especially the feedback about the surveys helped us improving them in each iteration.
One disadvantage of such a broad user survey is the high number of incomplete responses.
However, after the first survey, we managed to decrease the proportion of incomplete responses by optimizing the question structure.

\begin{table}
    \centering
    \begin{tabularx}{.75\linewidth}{ccccccc}
        \toprule
        & & \multicolumn{2}{c}{\small \textbf{Responses}} & \multicolumn{3}{c}{\small\textbf{Time (in minutes)}}\\
        \# & Execution & Total & Full & 25\% & 50\% & 75\%\\
        \midrule
        1 & 2020-10 & 84 & 37 & 41 & 75 & 126 \\
        2 & 2021-04 & 98 & 44 & 17 & 35 & 57 \\
        3 & 2021-10 & 85 & 54 & 24 & 38 & 57 \\
        \bottomrule
    \end{tabularx}
    \caption{Metadata of the conducted surveys including the month of execution, the number of total resp.\ full responses, and the average resp.\ median time per full response.}\label{table:survey:overview}
\end{table}

\subsubsection{Survey Participants}
Gathering participants for user surveys with a highly specific topic like decompilation or reverse engineering is not particularly easy.
Certainly, the first choice for participants are professional reverse engineers or people who are professionally active in malware analysis.
Unfortunately, there are neither plenty professionals in these areas nor are they available to take part in time-consuming studies resulting in fewer participants compared to less-technical surveys.

However, it is arguable that people with a solid understanding of the C programming language can participate in our survey as well.
First, most decompiler outputs are syntactically very similar to C code which is comprehensible by programmers, computer scientists, and others.
Second, comprehension and preference for the format of C-like source code does not require reversing skills.
Finally, the ultimate goal is to generate output also non-experts can comprehend because learning reversing and assembly is very difficult and time-consuming~\cite{mantovani2022remind}.
Consequently, we did not limit our survey to participants with experience in reverse engineering and were pleasantly surprised by the extensive participation.

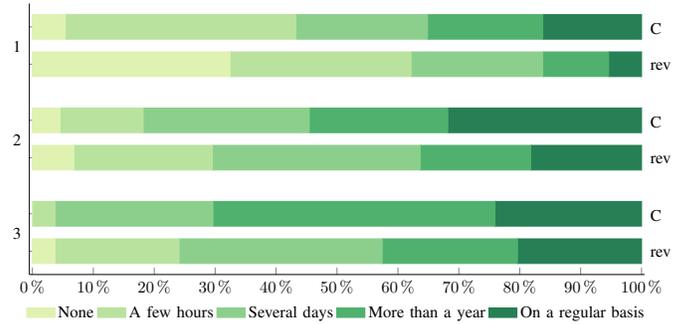
\begin{figure}[tb]
    \begin{adjustbox}{width=\linewidth}%
        \centering
    \begin{tikzpicture}
        \begin{axis}[
            xbar stacked,
            legend style={
                legend columns=5,
                at={(xticklabel cs:0.5)},
                anchor=north,
                draw=none
            },
            ytick=data,
            axis y line*=none,
            axis x line*=bottom,
            height = 5cm,
            xtick={0,10,20,30, 40, 50, 60, 70, 80, 90, 100},
            xticklabel=\pgfmathprintnumber{\tick}\,$\%$,
            ytick = data,
            width=.80\textwidth,
            bar width=6mm,
            yticklabels=\empty,
            ylabel style={at={(current axis.left of origin)},xshift=15mm},
            xmin=0,
            xmax=100,
            area legend,
            y=8mm,
            enlarge y limits={abs=0.625},
            enlarge x limits={abs=0.5},
            bar width=15pt,
            clip=false
        ]
        \addplot[colorg1!85,fill=colorg1!85] coordinates
        {(5.405405405405405, 6.5) (32.432432432432435, 5.5) (4.545454545454546, 4) (6.818181818181818, 3) (0.0, 1.5) (3.7037037037037037, 0.5)};
        \addplot[colorg2!85,fill=colorg2!85] coordinates
        {(37.83783783783784, 6.5) (29.72972972972973, 5.5) (13.636363636363637, 4) (22.727272727272727, 3) (3.7037037037037037, 1.5) (20.37037037037037, 0.5)};
        \addplot[colorg3!85,fill=colorg3!85] coordinates
        {(21.62162162162162, 6.5) (21.62162162162162, 5.5) (27.272727272727273, 4) (34.09090909090909, 3) (25.925925925925927, 1.5) (33.333333333333336, 0.5)};
        \addplot[colorg4!85,fill=colorg4!85] coordinates
        {(18.91891891891892, 6.5) (10.81081081081081, 5.5) (22.727272727272727, 4) (18.181818181818183, 3) (46.2962962962963, 1.5) (22.22222222222222, 0.5)};
        \addplot[colorg5!85,fill=colorg5!85] coordinates
        {(16.216216216216218, 6.5) (5.405405405405405, 5.5) (31.818181818181817, 4) (18.181818181818183, 3) (24.074074074074073, 1.5) (20.37037037037037, 0.5)};
        \legend{None,A few hours,Several days,More than a year, On a regular basis}
        \node at (xticklabel cs:-0.02,-154) {1};
        \node at (xticklabel cs:-0.02,-97) {2};
        \node at (xticklabel cs:-0.02,-40) {3};

        \node[anchor=west] at (xticklabel cs:1,-165) {C};
        \node[anchor=west] at (xticklabel cs:1,-142) {rev};
        \node[anchor=west] at (xticklabel cs:1,-108) {C};
        \node[anchor=west] at (xticklabel cs:1,-85) {rev};
        \node[anchor=west] at (xticklabel cs:1,-51) {C};
        \node[anchor=west] at (xticklabel cs:1,-28) {rev};

        \end{axis}
    \end{tikzpicture}
    \end{adjustbox}
    \caption{The participants' reversing and C experience across all user surveys. For the first survey (ranking from~1 to~10), we merged~1+2 to \emph{None}, 3+4 to \emph{A few hours}, and so on.}
    \label{fig:eval:experience}
\end{figure}

In~\Cref{fig:eval:experience} one can see that most participants have experience with C according to their self-assessment (see~\Cref{sec:eval:limitations}).
In the second and third survey roughly~60\% of the participants stated they had C-experience.
As expected, the number of participants with reversing skills is comparatively low.
Nevertheless, we gathered more than~40\% of participants with allegedly substantial reversing skills in the final survey.
The contribution of the skill levels is not surprising since our participants mainly consisted of students who completed a lecture in malware or program analysis, colleagues with experience in malware analysis or at least with C coding skills, and some professional reverse engineers.

\subsubsection{Survey Platform}
We opted to use an online platform to conduct our surveys to not be limited to geographically close participants, to avoid time constraints, and to ensure anonymous data collection.
More specifically, we used a privately hosted instance of LimeSurvey~\cite{engard2009limesurvey}.
At the beginning of each survey, we sent a survey link to potential participants.
In total, each survey was active for two weeks, allowing the participants to find time to complete the survey questionnaire.

Due to the page limit, we will only discuss a subset of the survey questions and results in this paper.
\Cref{sec:appendix} contains a few more details on some of the results.
Additionally, we separately published the complete questionnaires, results, and samples of all surveys on GitHub~\cite{survey-results}.

\subsection{Survey 1: Limitations of Dream}
One main goal of the first user survey was to evaluate how our re-implementation of the \dream approach performs against the commercial and open-source state-of-the-art.
We decided to use the Ackermann function (\Cref{fig:eval:ackermann}) compiled with GCC~10.0.1 since it is a non-trivial arithmetic function featuring non-primitive recursion.
Then, we decompiled the Ackermann function with dewolf (\dream re-implementation), Ghidra~9.1.2., and Hex-Rays~7.5.\ SP2, and we randomly assigned each participant one output.
Overall, we collected data about the comprehension and subjective trends regarding the presentation of the three different outputs.

\begin{figure}[tb]
    \begin{minted}[autogobble, bgcolor=my-bg,linenos,numbersep=4pt, breaklines,fontsize=\scriptsize]{c}
    int A(int m, int n) {
        if (!m) return n + 1;
        if (!n) return A(m - 1, 1);
        return A(m-1,A(m,n-1)); }
    \end{minted}
    \vspace{-0.6cm}
    \caption{Source code snippet for the first user survey. The function \mintinline[bgcolor=my-bg,linenos,numbersep=4pt, breaklines]{c}{A(...)} implements the Ackermann function.}
    \label{fig:eval:ackermann}
\end{figure}

\begin{figure}[tb]
    \begin{adjustbox}{width=\linewidth}%
        \centering
    \begin{tikzpicture}
        \begin{axis}[
            ybar stacked,
            y label style={at={(axis description cs:0.1,.5)},anchor=south},height = 4cm, width= 0.5\textwidth, major x tick style = transparent,
            ymin=0, ymajorgrids = true, xtick = data, ytick={0, 20, 40, 60, 80, 100},
            bar width=10pt, yticklabel=\pgfmathprintnumber{\tick}\,$\%$,
            xmin=dewolf1, xmax=Ghidra3, enlarge x limits={0.1},
            legend style={at={(0.3,0.3)},
              anchor=north,legend columns=-1},
            symbolic x coords={dewolf1, IDA1, Ghidra1, , dewolf2, IDA2, Ghidra2, , dewolf3, IDA3, Ghidra3},
            xticklabels={dewolf, IDA, Ghidra, dewolf, IDA, Ghidra, dewolf, IDA, Ghidra},
            x tick label style={rotate=60,anchor=east},
            clip=false
            ]
        \addplot+[ybar, color=colorg5!85] plot coordinates {(dewolf1, 93.33333333333333) (IDA1, 84.61538461538461) (Ghidra1, 77.77777777777777) (dewolf2, 86.66666666666667) (IDA2, 46.15384615384615) (Ghidra2, 55.55555555555556) (dewolf3, 73.33333333333333) (IDA3, 61.53846153846154) (Ghidra3, 66.66666666666667)};
        \addplot+[ybar, color=colorg3!85] plot coordinates {(dewolf1, 0.0) (IDA1, 15.384615384615385) (Ghidra1, 11.11111111111111) (dewolf2, 0.0) (IDA2, 38.46153846153846) (Ghidra2, 11.11111111111111) (dewolf3, 26.666666666666668) (IDA3, 38.46153846153846) (Ghidra3, 33.333333333333336)};
        \addplot+[ybar, color=colorg1!85] plot coordinates {(dewolf1, 6.666666666666667) (IDA1, 0.0) (Ghidra1, 11.11111111111111) (dewolf2, 13.333333333333334) (IDA2, 15.384615384615385) (Ghidra2, 33.333333333333336) (dewolf3, 0.0) (IDA3, 0.0) (Ghidra3, 0.0)};
        \legend{Correct, Wrong, None}
        \node at (xticklabel cs:0.17,-123) {Question 1};
        \node at (xticklabel cs:0.5,-123) {Question 2};
        \node at (xticklabel cs:0.83,-123) {Question 3};
        \end{axis}
        \end{tikzpicture}
    \end{adjustbox}
    \caption{Comprehension results of the Ackermann function. The questions are included in~\Cref{appendix:survey1}.}
    \label{fig:eval:ackermann_results}
\end{figure}
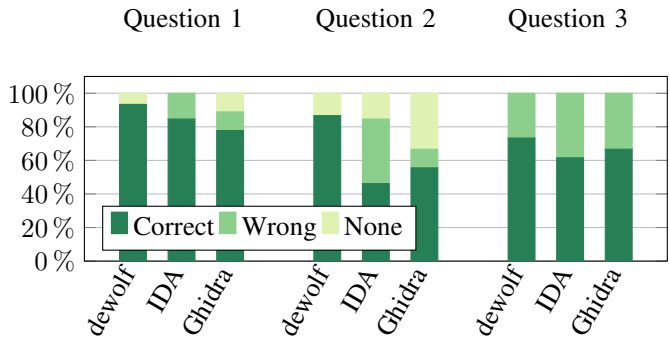

To verify how well the participants understood the given function and to validate the participant's self-assessment, we asked which values the function returns for two different sets of parameters and which parameter combinations result in the most recursive calls.
\Cref{fig:eval:ackermann_results} shows the comprehension results divided by the decompiler that generated the output for the respective participant.
Additionally, we asked some questions about the output quality.
Since we identified no grave differences between dewolf and the other decompilers in comprehension and quality, we felt confident that dewolf produces competitive results.

We used the second part of the survey to compare small decompiled samples of all three decompilers.
Therefore, each participant was presented with all three outputs side-by-side and rated them concerning various criteria like different control-flow structures.
One goal was to detect possible improvements for the re-implementation of \dream, and another goal was to gain an insight into user preferences.

Overall, our decompiler achieves results only slightly diverging from the state-of-the-art, except for compiler idioms, which were handled neither in the initial implementation nor the \dream approach.
Furthermore, the survey indicates that even the re-implementation of \dream can still compete with state-of-the-art decompilers.
Consequently, \dream provides a great starting point for additional research.

In the third part of the survey, we tried to find general user preferences regarding decompiler output.
On the one hand, there are many code constructs or output formats where researchers and users have a clear consensus.
For example, most would agree that a high nesting depth or representing an integer value as a binary number is generally undesirable.
On the other hand, there is a variety of cases where seemingly no clear preference exists such as the usage of gotos.
Therefore, we selected code constructs to understand user preferences regarding different aspects.
Further, we investigated the construct's effects on comprehension from a user's perspective.
Additionally, we validated whether ideas that originated from related work or previous research are worth further investigation.
The topics we asked participants include but are not limited to preferences regarding program structure like switch versus if-else, value and expression propagation, the use of goto, duplicate code, loop types, the display of various condition types, and common subexpression elimination.

It turned out that many participants were particularly dissatisfied with seemingly small details that may still require elaborated approaches, for example, the representation of constants or the restructuring of (for-)loops.
Moreover, the fact that modern decompilers do not handle all compiler idioms turned out to be a substantial issue.
Consequently, we decided to tackle issues criticized by participants regarding the output of our re-implementation of \dream.
Furthermore, we also approached two more complex and particularly substantial issues mentioned by participants, namely the complexity of expressions and conditions and the number of variables and their names.
To address these problems, we developed a custom logic engine (\Cref{sec::logic}) and focused on out-of-SSA as well as on handling variable copies.

Another aspect frequently mentioned by participants was the lack of configurability.
Indeed, users may have contrasting preferences and can benefit from diverse configuration options.
Additionally, preferences often diverge depending on the given sample or function.
Hence, it is no surprise that most participants are disappointed with the extent to which some algorithms of state-of-the-art decompilers are configurable.
More specifically, they would appreciate tweaking decompiler settings to fit their individual preferences or the characteristics of a given function.
Consequently, in developing dewolf and new approaches, we considered numerous possibilities for new configurations.
Therefore, many algorithms in dewolf are highly configurable by the user.

\subsection{Survey 2: Comprehension}
The goal of the second survey was to evaluate the comprehensiveness of dewolf's output.
More precisely, to assess the applicability of dewolf, we validated whether participants can use dewolfs output to successfully comprehend a realistic malware function.
We decided that an adequate malware component to be analyzed in our survey would be a domain generation algorithm (DGA).
DGAs implement a kind of domain flux designed to improve the resilience of network communication utilized by modern malware to avoid a single-point-of-failure during network communication by not using hardcoded addresses for the network infrastructure of the malware~\cite{plohmann2016comprehensive}.
As a consequence, during malware analysis, analysts are often interested in the \emph{detailed} implementation of DGAs to obtain domains of interest, in contrast to e.g., crypto algorithms where no complete understanding may be required once identified.

Unfortunately, we only have a limited number of survey participants and each only agrees to spend a limited amount of time for the surveys.
In addition, real-world malware samples are exponentially increasing in size as well as complexity~\cite{calleja2016look}, and DGAs regularly are implemented across multiple functions and may have dependencies which are both hard to represent properly in a user survey.
Consquently, we had to decide between a real-world DGA that potentially would lead to less or even none completed survey participations and a synthetic DGA which we opted for eventually.
To make the DGA as realistic as possible, we studied numerous real-world DGAs using DGArchive~\cite{plohmann2016comprehensive} and programmed a very similar yet slightly less complex DGA in a single function that still generates realistic domains as a real-world DGA would.
\Cref{fig:eval:dga} shows the DGA used for this survey.

\begin{figure}[tb]
    \begin{minted}[autogobble, bgcolor=my-bg,linenos,numbersep=4pt, breaklines,fontsize=\scriptsize]{c}
    char* dga() {
      unsigned char* domain; char seed;
      domain = malloc(sizeof(SYSTEMTIME));
      seed = (char)GetTickCount();
      GetSystemTime((LPSYSTEMTIME)domain);
      for(char i = 0; i < DOMAIN_LENGTH; i++) {
        domain[i]=((unsigned char)(domain[i]^seed)%24)+97;}
      int* end = domain + DOMAIN_LENGTH;
      end[0] = "\x2e\x63\x6f\x6d"; end[1] = 0;
      switch(seed % 8){
        case 7: case 5:
          end[0] ^= "\x00\x11\x1a\x6d"; break;
        case 1: case 6:
          end[0] ^= "\x00\x17\x00\x6d"; break;
        case 2:
          end[0] ^= "\x00\x0d\x0a\x19"; break;
        default: break; }
      return domain; }
    \end{minted}
    \vspace{-0.6cm}
    \caption{The artificial domain generation algorithm of the second survey which produces domains with~8 random characters excluding the letter \texttt{z} with four different TLDs.}
    \label{fig:eval:dga}
\end{figure}

Each participant received dewolf's output of the given DGA, compiled with VS2015, depicted in~\Cref{fig:eval:dga:dewolf}.
Then, we asked the participants about the purpose of the function to assess whether the participants determine its functionality, i.e., that it is a DGA.
In total, more than~60\% of the participants were able to identify the purpose of the function correctly.
Noticeably, the majority of participants that gave an incorrect answer had either no C- respectively reversing skills or rushed through the survey in less than~15 minutes.
Since we continued with more in-detail questions involving the structure and form of the generated domains, we explained the function's purpose to all participants after the first question.
More specifically, we asked about the domain length, character set, and top-level domains, allowing us to establish which parts of the algorithm were understandable by the participants.
Most participants were able to identify the used TLDs and correctly classified \emph{.com} as the most used TLD.
Finally, we showed the participants five domains and asked whether the given algorithm could potentially generate these domains.
Again, the majority of the participants answered those questions correctly.
The results of the comprehension questions are summarized in~\Cref{fig:eval:dga_comprehension} whereas all questions are contained in~\Cref{appendix:survey2}.

\begin{figure}[tb]
    \begin{adjustbox}{width=\linewidth}%
        \centering
    \begin{tikzpicture}
        \begin{axis}[y label style={at={(axis description cs:0.1,.5)},anchor=south},xlabel={Number of Correct Answers}, ylabel={Number of Participants},ybar=4*\pgflinewidth,bar width=3pt, height = 6cm, width  = 0.55\textwidth, symbolic x coords={0, 1, 2, 3, 4, 5, 6, 7, 8, 9, 10, 11}, major x tick style = transparent,ymin=0,ymajorgrids = true, xtick = data, legend style={cells={anchor=west}, legend pos=north west},xmin=0, xmax=11, enlarge x limits={0.05}, ytick={0, 2, 4, 6, 8, 10, 12}]
            \addplot [bar shift=-6pt, style={colorg1,fill=colorg1}]
            coordinates{(0, 0)(1, 0)(2, 2)(3, 0)(4, 0)(5, 0)(6, 0)(7, 0)(8, 0)(9, 0)(10, 0)(11, 0)};
            \addplot[bar shift=-3pt, style={colorg2,fill=colorg2}]
            coordinates{(0, 0)(1, 1)(2, 0)(3, 0)(4, 2)(5, 0)(6, 1)(7, 0)(8, 0)(9, 0)(10, 1)(11, 1)};
            \addplot[bar shift=0pt, style={colorg3,fill=colorg3}]
            coordinates{(0, 2)(1, 0)(2, 0)(3, 0)(4, 2)(5, 0)(6, 1)(7, 0)(8, 1)(9, 1)(10, 2)(11, 3)};
            \addplot[bar shift=3pt, style={colorg4,fill=colorg4}]
            coordinates{(0, 0)(1, 1)(2, 0)(3, 0)(4, 0)(5, 0)(6, 1)(7, 0)(8, 1)(9, 3)(10, 1)(11, 3)};
            \addplot[bar shift=6pt, style={colorg5,fill=colorg5}]
            coordinates{(0, 2)(1, 0)(2, 0)(3, 0)(4, 0)(5, 0)(6, 1)(7, 0)(8, 0)(9, 2)(10, 4)(11, 5)};
            \begin{pgfonlayer}{background}
            \addplot[bar shift=0pt, style={my-bg!70!black,fill=my-bg!80!black}, bar width=15pt] coordinates{(0, 4)(1, 2) (2, 2) (3, 0) (4, 4) (5, 0) (6, 4) (7, 0) (8, 2) (9, 6) (10, 8)(11, 12)};
            \legend{None, A few hours, Several days, More than a year, On a regular basis, All skill levels}
            \end{pgfonlayer}
        \end{axis}
    \end{tikzpicture}
    \end{adjustbox}
    \caption{DGA Comprehension results of the second survey for all participants and split by C-skill level.}
    \label{fig:eval:dga_comprehension}
\end{figure}
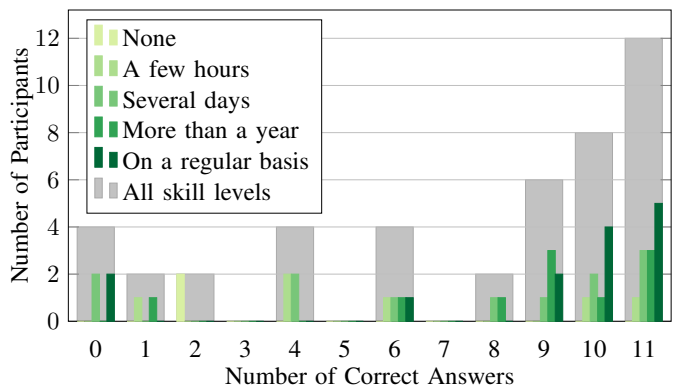

Overall, slightly more than~70\% of all participants answered at least half of the questions correctly.
Since most of the questions were not trivial, we are delighted by these results.
Further,~\Cref{fig:eval:dga_comprehension} shows that C skills seem to have an impact on the answer score of the participants.
This impact is not surprising and strengthens the result even more because most incorrect answers were seemingly not caused by dewolf's output but by the lack of expertise.

Finally, the participants compared dewolf's output of the DGA with a few other decompilers.
The goal was to collect valuable feedback, user opinions about the different decompiler outputs, and possible improvements.
The results indicate that for the given sample, Hex-Rays, Ghidra, and dewolf produce an eminently better output than the others which we consequently did not consider in the final survey.
Similar to the first survey, this feedback significantly contributed to the approach introduced in~\Cref{sec:approach}.
Overall, the second survey demonstrates that dewolf produces output that allows analyzing malware successfully.
Furthermore, it appears that adequate C skills can be advantageous but are not crucial.

\subsection{Survey 3: Comparison}
The primary goal of the final user survey was to compare our approach introduced in~\Cref{sec:approach} (dewolf) with other decompilers.
Due to the instability of the \dream implementation, we only compared dewolf to Ghidra~10.0.3.\ and Hex-Rays~7.61.\ SP1.
Besides, we already compared our initial re-implementation of \dream to both Hex-Rays and Ghidra in the first user survey.
We opted against comparing our approach to additional approaches due to the survey limitations discussed in~\Cref{sec:eval:limitations}.
Additionally, at the time of conducting the third survey, neither the Pseudo-C output from Binary Ninja nor revng-c were publicly released.
The output from other decompilers we included in the second survey turned out to produce output so significantly less readable than Hex-Rays and Ghidra that we chose to not consider them in the final user survey (see~\cite{survey-results}).

For the comparison, we constructed a sample that contains as many different difficult decompilation aspects as possible while retaining a manage-able size.
We compiled the sample function from~\Cref{fig:eval:sample_survey3} with GCC~10.3.1 and decompiled the resulting sample with Hex-Rays and Ghidra (\Cref{fig:appendix:survey3:outputs}).
In contrast to the first survey, each participant was presented with all three samples in random order and marked with pseudonyms.
Then, each participant ranked them from most to least favorable (\Cref{fig:eval:ranking}).
The results indicate that dewolf was the most favored for the considered function.
Also, Hex-Rays was still significantly ahead of Ghidra.
This result clearly underlines the positive impact of all introduced approaches, although each individually being relatively simple, at least on the participants' perception.

\begin{figure}[tb]
    \begin{minted}[autogobble, bgcolor=my-bg,linenos,numbersep=4pt, breaklines,escapeinside=||,fontsize=\scriptsize]{c}
    int convert_binary_to_hex() {
        long long binary; char hex[65] = "";
        int remainder;
        printf("Enter any binary number: "); scanf("%lld", &binary);
        while(binary > 0) {
            remainder = binary % 10000;
            switch(remainder) {
                case 0: strcat(hex, "0"); break;
                case 1: strcat(hex, "1"); break; /* ... */
    \end{minted}
    \vspace{-0.9cm}
    \begin{minted}[autogobble, bgcolor=my-bg,linenos,numbersep=4pt, breaklines,escapeinside=||,fontsize=\scriptsize]{c}
    |\setcounter{FancyVerbLine}{22}|            case 1110: strcat(hex, "E"); break;
                case 1111: strcat(hex, "F"); break; }
            binary /= 10000; }
        printf("Binary number: %lld\n", binary);
        printf("Hexadecimal number: %s", hex);
        return 0; }
    \end{minted}
    \vspace{-0.6cm}
    \caption{Code Snippet for the third user survey to generate the hex representation of a given binary number.}
    \label{fig:eval:sample_survey3}
\end{figure}

All in all, the participants positively mentioned several aspects of dewolf's output.
For instance,~44 participants praised the switch structure we reconstructed in contrast to a higher nesting depth produced by the other two decompilers.
Another convincing aspect was the lack of unnecessary casts in dewolf's output.
However, the participants also disliked some details about dewolf's output.
For example, dewolf was incapable of correctly detecting the array, whereas Hex-Rays was.
To clarify, although dewolf was ranked first in this sample, we do not claim that dewolf is superior in general.
Nevertheless, we want to point out that dewolf is capable of applying restructuring techniques not supported by the other decompilers and exceeds their output in certain properties.

\begin{figure}[tb]
    \centering
    \begin{adjustbox}{width=0.95\linewidth}%
    \centering
    \begin{tikzpicture}
        [
        pie chart,
        slice type={dewolf}{colorg5!75},
        slice type={Hex-Rays}{colorg3!75},
        slice type={Ghidra}{colorg1!75},
        pie values/.style={font={\small}},
        scale=1.5
        ]

        \pie{Rank 1}{83/dewolf,
            13/Hex-Rays,
            4/Ghidra}
        \pie[xshift=2.2cm,values of coltello/.style={pos=1.1}]%
        {Rank 2}{16/dewolf,
        70/Hex-Rays,
        14/Ghidra}
        \pie[xshift=4.4cm,values of caffe/.style={pos=1.1}]%
        {Rank 3}{1/dewolf,
        18/Hex-Rays,
        81/Ghidra}

        \legend[anchor=south,xshift=-0.85cm,yshift=-1.35cm]{{dewolf}/dewolf, {Hex-Rays}/Hex-Rays, {Ghidra}/Ghidra}

    \end{tikzpicture}
    \end{adjustbox}

    \caption{Ranking of the decompilers for the function in~\Cref{fig:eval:sample_survey3}. A total of~53 participants ranked dewolf best.}
    \label{fig:eval:ranking}
\end{figure}
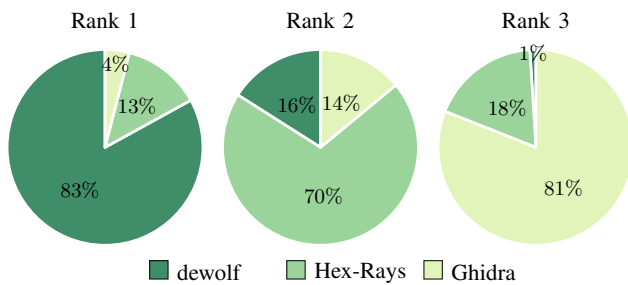

In the second part of this survey, we (again) asked the participants some questions regarding their preferences for decompiler output.
The idea behind this part was to question the generally taken assumption that reverse engineers want highly accurate decompilation output, i.e., an output similar to the assembly structure.
Unsurprisingly, many decompilation approaches and commercial decompilers aim to achieve this goal~\cite{verbeek2020sound,brumley2013native}.
Nevertheless, some approaches like \dream may already cause functions being restructured slightly different from their original appearance to some extent.
Overall, our impression is that most approaches try to generate output very similar to the assembly structure.

We decided to scratch the surface of this unwritten rule by including a few questions to verify whether survey participants would accept substantial differences to the assembly structure in favor of increasing the decompiler output's readability.
For instance, the participants had to choose between three equivalent reconstruction options for the same function.
The function was restructured by either having a more complicated structure, containing various levels of duplicated code, or an alternative version where a section was extracted into a new artificial function.
In total, we evaluated three restructuring approaches for different scenarios to obtain an overall impression.
\Cref{appendix:fig:extraction} contains an example of this question~type.

As expected, some participants complained that a decompiler should not alter the structure of the considered program imposed by the disassembly.
However, depending on the scenario, the majority preferred the alternative version even though being clearly different from the assembly structure.
Furthermore, some participants specifically praised the better readability translating to less required analysis time and possibly better analysis results.

\subsection{Survey Limitations and Threats to Validity} \label{sec:eval:limitations}
The evaluation of a decompilation approach with user surveys has certain limitations, and consequently has our evaluation.
Nevertheless, we will point out in this section why user surveys are favorable when evaluating readability improvements despite their limitations.
The most notable limitation is that only a small number of people are willing and able to participate in a user survey related to reversing.
Especially the number of professional reversers is rather small and becomes even smaller when only considering the ones willing to spend time on a user survey.
Additionally, recent research indicates that even popular paid services claiming to provide qualified survey participants can and should not be used to overcome this limitation~\cite{danilova2021you}.
Consequently, like previous studies with a similar topic~\cite{yakdan2015no,mantovani2022remind}, we have to work with a relatively limited number of participants compared to less-technical user surveys.

Next, the limited number of participants also influences the survey extent as each participant is only willing to spend a certain amount of time.
The first user survey shows that most participants would not spend significantly more than 30min to complete a survey.
Even the samples we utilized in our surveys resulted in noticeably longer participation times (see~\Cref{table:survey:overview}) and already caused issues with participants not completing the survey.
Thus, using real-world malware samples which are far more complex, individually evaluating the impact of all improvements, or comparing our approach against more than one or two other approaches is unfeasible.
Therefore, we decided to evaluate the combined impact of all introduced improvements of dewolf using real-world-like samples.
To tweak the difficulty and participation time, we had student assistants do test runs of each survey.

The last limitation of our survey design is the lack of an appropriate assessment of participants' skills for each survey.
In the first survey, we included both a self-assessment and an assessment based on questions about the Ackerman Function.
Since we did not observe any significant divergences between those two, we decided to only include self-assessment questions in the following surveys.
Having only the self-assessment saved valuable participants' time and allowed us to include additional content.
Consequently, when considering results split by the participant's skills, these are solely built on the provided self-assessment and should be tempered with caution.
Regardless, the survey results indicate that the participant's reversing skills are far less relevant compared to the capability of understanding code which is very helpful for comprehending a given function.

Despite all those limitations of user surveys, they are still the best way to evaluate approaches that improve readability for analysts.
Although researchers commonly use code metrics to assess the quality of their code-base, these are inadequate for our evaluation:
The first user study shows that no sound default configuration for decompiler outputs exists.
Consequently, it is impossible to define formal criteria for code quality because readability is merely subjective.
To summarize, regardless of the discussed limitations (see~\Cref{sec:eval:limitations}) user surveys are the most fitting method to evaluate decompilation approaches focusing on readability.

\subsection{Runtime Evaluation}
\begin{figure}[tb!]
    \linespread{0.0}
    \centering
    \begin{adjustbox}{width=\linewidth}%
        \centering
        \begin{tikzpicture}
            \begin{axis}
                [
                width=0.8\linewidth,
                xtick={-3,-2,-1,0,1,2,3},
                xmin=-3.9,
                xmax=3.1,
                xticklabels={%
                    {10\textsuperscript{-3}},
                    {10\textsuperscript{-2}},
                    {10\textsuperscript{-1}},
                    {10\textsuperscript{0}},
                    {10\textsuperscript{1}},
                    {10\textsuperscript{2}},
                    {10\textsuperscript{3}}
                },
                x tick label style={anchor=north,font=\scriptsize},
                xlabel={\scriptsize Time in seconds},
                xlabel style={yshift=.5em},
                height = 2.8cm,
                ytick={1,2},
                yticklabels={Binary\\ Ninja, dewolf},
                y tick label style={
                align=right,
                font=\scriptsize
                },
                cycle list={{colorg1!75},{colorg5!75}},
                boxplot/every median/.style={black,thick,solid},
                boxplot/every whisker/.style={black},
                every axis plot/.append style={fill,fill},
                ]
                \addplot+[
                boxplot prepared={
                    median=-2.442789928,
                    upper quartile=-2.089656697,
                    lower quartile=-2.739665967,
                    upper whisker=0.128181802,
                    lower whisker=-3.590910937
                },
                ] coordinates {};
                \addplot+[
                boxplot prepared={
                    median=0.759932917,
                    upper quartile=0.596197369,
                    lower quartile=0.907330231,
                    upper whisker=2.47671537,
                    lower whisker=0.461335389
                },
                ] coordinates {};
            \end{axis}
        \end{tikzpicture}
    \end{adjustbox}
    \caption{Boxplots visualizing the runtime when decompiling the samples in Coreutils with dewolf and Binary Ninja.}
    \label{fig:timing_results:boxplots}
\end{figure}
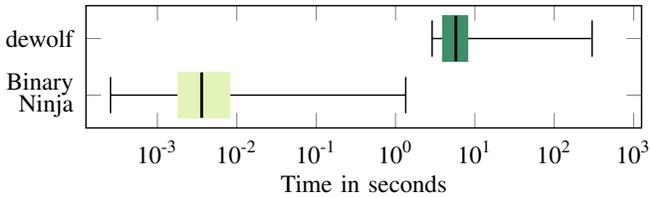

We heavily focus on readability improvements and consequently neither optimized our approach nor our prototype implementation in terms of speed.
While we want to leave a precise runtime evaluation of the presented approaches' subject for future work, we understand that a \textit{reasonable} runtime is an important aspect of decompilation approaches.
To demonstrate that even without speed optimizations or being implemented in native code, our Python prototype is sufficiently fast on real-world binaries, we decompiled all binaries in Coreutils on an i7-10850H with 12GBs of RAM and measured the decompilation times.
Given a timeout of 5 minutes, the dewolf prototype was able to decompile 94\% of all binaries and 75\% of them in less than 8 seconds.
\Cref{fig:timing_results:boxplots} shows the measured decompilation times for dewolf and also for Binary Ninja.
As presumed, the decompilation times of Binary Ninja are significantly faster.
Because all of Binary Ninja's analyses run at the start of dewolf to obtain the MLIL, they can be considered a lower bound for dewolf's runtime.

\subsection{Summary and Outlook}
We conclude that future research on decompilation should presumably not be limited to reconstructing a binary's assembly structure as accurately as possible.
Instead, to help human analysts, it may be highly beneficial to improve readability and comprehensibility by allowing more flexible output generation while maintaining semantic equivalence.
Our survey results indicate that users accept decompiler output with changed program structure to improve the readability.
Consequently, we have to overthink existing approaches that heavily avoid substantial changes to the program structure and instead take them actively into consideration.
Ultimately, our results establish multiple possibilities for decompilation, opening entirely new options for output formats.
Additionally, it encourages techniques that were previously hastily discarded due to the above reasons.

The results of our surveys also imply that configurability is deeply desired by many participants.
Because readability seems to be highly subjective, analysts can benefit from individual high-level representations influenced by decompiler parameters.
Besides, as the characteristics of functions vastly differ between samples, configuration options could be very useful to tune the decompiler for each given use-case.
In conclusion, configurability should be considered for future decompilation approaches.

Finally, we expect that the survey results may be helpful to other researchers in the field of decompilation, reverse engineering, or even topics dealing with comprehending source code.
In particular, it is possible to use some of the results as a baseline for further research.
We publish the complete questionnaires and participants' responses to allow further analyses and conclusions from others~\cite{survey-results}.

\section{Conclusion}
\label{sec::conclusion}
In this paper, we made three main contributions.
First, we conducted and discussed three user surveys providing valuable information for research in the field of reverse engineering and decompilation.
We publish the survey results allowing other researchers to use them in three ways:
(i) as a guideline of user preferences for decompiler output or code in general,
(ii) insights to the limitations on current approaches from a user perspective, and
(iii) as starting points for new research.
A key finding of our surveys is that the majority of participants do not necessarily favor decompilers producing output as close as possible to the structure implied by the assembly.
Instead, we discovered that users prefer more readable and comprehensive code even when it introduces substantial structural differences to the disassembly.
These results may have a broad impact on future research since existing approaches rarely oppose the disassembly.
Furthermore, the evaluation indicates that rather small details, like the representation of constants, can already have a positive impact in some cases.
Consequently, future decompilation approaches should allow user configuration for such details when possible.
This result raises the question of whether the existing metrics for decompiler evaluation apply to human analysis.

Based on the survey results, we introduced a new decompilation approach named dewolf with significant improvements over the previous research state-of-the-art.
In addition to countless small-scale improvements, we developed a custom logic engine and a novel and well-described out-of-SSA algorithm.
The evaluation indicates that dewolf is not only a considerable improvement over the previous state-of-the-art but is also suitable to comprehend and analyze malware.
Moreover, the output quality of dewolf noticeably exceeds the commercial and open-source state-of-the-art decompilers Hex-Rays and Ghidra for several samples, according to survey participants.

Lastly, we open-source our decompiler prototype dewolf to provide other researchers with a suitable base for their work.
Due to its stability and its modular design, dewolf is exceptionally well suited to integrate and evaluate new approaches.
Further, we hope that dewolf motivates and assists more research focusing on manual analysis to cope with the increasing security analysis load.

\bibliographystyle{IEEEtranS}
\bibliography{IEEEabrv,references}
\appendix \label{sec:appendix}
\subsection{User Survey I}
\label{appendix:survey1}
\Cref{fig:appendix:survey1:outputs} shows the decompiler outputs we generated with dewolf, at that time essentially being a re-implementation of \dream, Hex-Rays, and Ghidra.
To reduce the space consumption of the outputs, we deleted some linebreaks and inline selected conditions.
We showed each participant one of those random decompiler outputs.
The participants then had to answer a set of questions about its comprehensiveness and their judgment of the output quality on a Likert scale.
\Cref{fig:appendix:survey1:boxplots} shows the overall results for each question for all three considered decompilers indicating only small divergences in most categories.
Four questions expose divergence between them and their control questions.
Interestingly, they contain all questions of two categories: variables and restructuring.
Apparently, participants do not have a general preference on the number of variables, their typing, and restructuring.

\begin{figure}[bth]
    \begin{subfigure}[t]{\linewidth}
        \begin{minted}[autogobble, bgcolor=my-bg,linenos,numbersep=4pt, breaklines,fontsize=\scriptsize]{c}
        unsigned long A(int arg1, int arg2) {
          unsigned long var_0;
          if (arg1 == 0) { var_0 = arg2; }
          else {
            var_0 = arg2;
            while(true) {
              arg2 = (arg1 + -0x1) & 0xffffffff;
              if ((unsigned int)var_0 == 0) {
                if (arg2 == 0) { var_0 = 0x1; break; }
                arg1 = arg2; var_0 = 0x1; continue; }
              var_0 = A(arg1, ((unsigned int) var_0)+0xffffffff);
              if (arg2 == 0) { break; }
              arg1 = arg2; } }
          return ((unsigned int) var_0) + 0x1; }
        \end{minted}
        \vspace{-0.6cm}
        \subcaption{Output generated with our initial re-implementation of \dream.}
        \label{fig:appendix:survey1:dewolf}
    \end{subfigure}
    \begin{subfigure}[t]{\linewidth}
        \begin{minted}[autogobble, bgcolor=my-bg,linenos,numbersep=4pt, breaklines,fontsize=\scriptsize]{c}
        __int64 __fastcall A(__int64 arg_1, int arg_2){
          unsigned int var_0;
          if ( (_DWORD)arg_1 ) {
            do{
              while ( 1 ) {
                var_0 = arg_1 - 1;
                if ( !arg_2 ) { break; }
                arg_2 = A(arg_1, (unsigned int)(arg_2 - 1));
                arg_1 = var_0;
                if (!var_0) {return (unsigned int)(arg_2 + 1); } }
              arg_2 = 1; arg_1 = var_0;
            }while ( var_0 ); }
          return (unsigned int)(arg_2 + 1); }
        \end{minted}
        \vspace{-0.6cm}
        \subcaption{Output generated with Hex-Rays and IDA 7.5.200728 (SP2).}
        \label{fig:appendix:survey1:hexrays}
    \end{subfigure}
    \begin{subfigure}[t]{\linewidth}
        \begin{minted}[autogobble, bgcolor=my-bg,linenos,numbersep=4pt, breaklines,fontsize=\scriptsize]{c}
        ulong A(ulong arg_1,ulong arg_2){
          int var_0; uint var_1; arg_2 = arg_2 & 0xffffffff;
          var_0 = (int)arg_2; var_1 = (uint)arg_1;
          while (var_1 != 0) {
            while( true ) {
              var_1 = (int)arg_1 - 1;
              if ((int)arg_2 != 0) { break; }
              arg_2 = 1; var_0 = 1; arg_1 = (ulong)var_1;
              if (var_1 == 0) { goto Label_1; } }
            arg_2 = A(arg_1,(ulong)((int)arg_2 - 1));
            var_0 = (int)arg_2; arg_1 = (ulong)var_1; }
          Label_1: return (ulong)(var_0 + 1); }
        \end{minted}
        \vspace{-0.6cm}
        \subcaption{Output generated with Ghidra~9.1.2.}
        \label{fig:appendix:survey1:ghidra}
    \end{subfigure}
    \caption{Decompiler outputs for the function from~\Cref{fig:eval:ackermann} used for the first user survey.}
    \label{fig:appendix:survey1:outputs}
\end{figure}

\begin{figure}[tb!]
    \linespread{0.0}
    \centering
    \begin{adjustbox}{width=\linewidth}%
        \centering
        \begin{tikzpicture}
        \begin{axis}[
            width=0.7\linewidth,
            xtick={-3,-2,-1,0,1,2,3},
            xmin=-3.1,
            xmax=3.1,
            xticklabels={%
                {strongly disagree},
                {disagree},
                {weakly disagree},
                {unsure},
                {weakly agree},
                {agree},
                {strongly agree}
            },
            x tick label style={rotate=90,anchor=east,font=\scriptsize},
            y=0.22cm,
            ymin=-0.3,
            ymax=68.3,
            ytick={0,4,8,12,16,20,...,100},
            y tick label as interval,%
            yticklabels={%
                {The used control-flow structures are appropriate.},%
                {The control-flow is strangely restructured.},%
                {It was easy to understand the code.},%
                {It took much effort to understand the code.},%
                {It seems that there are no unused instructions.},%
                {There are too much unused instructions.},%
                {I think the code is correctly decompiled.},%
                {The decompiled code seems to be incorrect.},%
                {The conditions are too complex.},%
                {The code contains too many intermediate results.},%
                {The line length is too long},%
                {Many variables have useless copies.},%
                {There are no useless copies of variables.},%
                {The variable types seem to be reasonable.},%
                {Some variable types seem to be wrong.},%
                {There are no variables that only store unnecessary intermediate constants.},%
                {There are too many variables that just store intermediate constants.},%
            },
            y tick label style={
                text width=2.8cm,
                align=right,
                font=\scriptsize
            },
            cycle list={{colorg1!85},{colorg3!85},{colorg5!85}},
            boxplot/every median/.style={black,thick,solid},
            boxplot/every whisker/.style={black},
            every axis plot/.append style={fill,fill},
            legend style={at={(-0.3cm,-0.3cm)},anchor=north east,font=\scriptsize},
            legend entries={dewolf, Hex-Rays, Ghidra},
            legend image code/.code={
                \fill [#1] (0cm,-0.1cm) rectangle (0.6cm,0.1cm);
            },
            ]

            \addplot+[
            boxplot prepared={
            draw position=1,
            median=1.0,
            upper quartile=2.0,
            lower quartile=0.0,
            upper whisker=3,
            lower whisker=-3
            },
            ] coordinates {};

            \addplot+[
            boxplot prepared={
            draw position=2,
            median=1.0,
            upper quartile=1.75,
            lower quartile=-1.0,
            upper whisker=3,
            lower whisker=-3
            },
            ] coordinates {};

            \addplot+[
            boxplot prepared={
            draw position=3,
            median=1.0,
            upper quartile=1.0,
            lower quartile=0.25,
            upper whisker=1,
            lower whisker=-2
            },
            ] coordinates {};

            \addplot+[
            boxplot prepared={
            draw position=5,
            median=0.0,
            upper quartile=1.0,
            lower quartile=-2.0,
            upper whisker=3,
            lower whisker=-2
            },
            ] coordinates {};

            \addplot+[
            boxplot prepared={
            draw position=6,
            median=1.0,
            upper quartile=1.75,
            lower quartile=-1.0,
            upper whisker=3,
            lower whisker=-2
            },
            ] coordinates {};

            \addplot+[
            boxplot prepared={
            draw position=7,
            median=1.0,
            upper quartile=2.0,
            lower quartile=-1.0,
            upper whisker=2,
            lower whisker=-2
            },
            ] coordinates {};

            \addplot+[
            boxplot prepared={
            draw position=9,
            median=0.0,
            upper quartile=1.0,
            lower quartile=-2.0,
            upper whisker=2,
            lower whisker=-3
            },
            ] coordinates {};

            \addplot+[
            boxplot prepared={
            draw position=10,
            median=1.0,
            upper quartile=1.0,
            lower quartile=-2.0,
            upper whisker=3,
            lower whisker=-2
            },
            ] coordinates {};

            \addplot+[
            boxplot prepared={
            draw position=11,
            median=-1.0,
            upper quartile=0.0,
            lower quartile=-1.5,
            upper whisker=1,
            lower whisker=-3
            },
            ] coordinates {};

            \addplot+[
            boxplot prepared={
            draw position=13,
            median=1.0,
            upper quartile=2.0,
            lower quartile=-1.0,
            upper whisker=3,
            lower whisker=-2
            },
            ] coordinates {};

            \addplot+[
            boxplot prepared={
            draw position=14,
            median=1.0,
            upper quartile=2.0,
            lower quartile=-1.0,
            upper whisker=3,
            lower whisker=-2
            },
            ] coordinates {};

            \addplot+[
            boxplot prepared={
            draw position=15,
            median=2.0,
            upper quartile=2.0,
            lower quartile=1.0,
            upper whisker=3,
            lower whisker=-1
            },
            ] coordinates {};

            \addplot+[
            boxplot prepared={
            draw position=17,
            median=1.0,
            upper quartile=2.0,
            lower quartile=0.0,
            upper whisker=3,
            lower whisker=-2
            },
            ] coordinates {};

            \addplot+[
            boxplot prepared={
            draw position=18,
            median=1.5,
            upper quartile=2.25,
            lower quartile=1.0,
            upper whisker=3,
            lower whisker=-1
            },
            ] coordinates {};

            \addplot+[
            boxplot prepared={
            draw position=19,
            median=1.5,
            upper quartile=2.0,
            lower quartile=0.5,
            upper whisker=3,
            lower whisker=-2
            },
            ] coordinates {};

            \addplot+[
            boxplot prepared={
            draw position=21,
            median=-1.0,
            upper quartile=-1.0,
            lower quartile=-2.0,
            upper whisker=3,
            lower whisker=-3
            },
            ] coordinates {};

            \addplot+[
            boxplot prepared={
            draw position=22,
            median=-1.0,
            upper quartile=-1.0,
            lower quartile=-2.0,
            upper whisker=3,
            lower whisker=-3
            },
            ] coordinates {};

            \addplot+[
            boxplot prepared={
            draw position=23,
            median=-1.0,
            upper quartile=-0.25,
            lower quartile=-1.25,
            upper whisker=2,
            lower whisker=-3
            },
            ] coordinates {};

            \addplot+[
            boxplot prepared={
            draw position=25,
            median=0.0,
            upper quartile=2.0,
            lower quartile=0.0,
            upper whisker=2,
            lower whisker=-2
            },
            ] coordinates {};

            \addplot+[
            boxplot prepared={
            draw position=26,
            median=1.0,
            upper quartile=2.0,
            lower quartile=1.0,
            upper whisker=3,
            lower whisker=-3
            },
            ] coordinates {};

            \addplot+[
            boxplot prepared={
            draw position=27,
            median=1.0,
            upper quartile=1.0,
            lower quartile=1.0,
            upper whisker=2,
            lower whisker=-2
            },
            ] coordinates {};

            \addplot+[
            boxplot prepared={
            draw position=29,
            median=0.0,
            upper quartile=0.0,
            lower quartile=-1.0,
            upper whisker=3,
            lower whisker=-2
            },
            ] coordinates {};

            \addplot+[
            boxplot prepared={
            draw position=30,
            median=-1.0,
            upper quartile=-1.0,
            lower quartile=-2.0,
            upper whisker=3,
            lower whisker=-2
            },
            ] coordinates {};

            \addplot+[
            boxplot prepared={
            draw position=31,
            median=-1.0,
            upper quartile=-0.5,
            lower quartile=-1.0,
            upper whisker=1,
            lower whisker=-1
            },
            ] coordinates {};

            \addplot+[
            boxplot prepared={
            draw position=33,
            median=-2.0,
            upper quartile=-1.0,
            lower quartile=-2.0,
            upper whisker=2,
            lower whisker=-3
            },
            ] coordinates {};

            \addplot+[
            boxplot prepared={
            draw position=34,
            median=-1.5,
            upper quartile=-1.0,
            lower quartile=-2.25,
            upper whisker=3,
            lower whisker=-3
            },
            ] coordinates {};

            \addplot+[
            boxplot prepared={
            draw position=35,
            median=-2.0,
            upper quartile=-1.0,
            lower quartile=-2.0,
            upper whisker=1,
            lower whisker=-2
            },
            ] coordinates {};

            \addplot+[
            boxplot prepared={
            draw position=37,
            median=0.0,
            upper quartile=1.0,
            lower quartile=-1.0,
            upper whisker=2,
            lower whisker=-2
            },
            ] coordinates {};

            \addplot+[
            boxplot prepared={
            draw position=38,
            median=-1.0,
            upper quartile=1.0,
            lower quartile=-2.0,
            upper whisker=3,
            lower whisker=-2
            },
            ] coordinates {};

            \addplot+[
            boxplot prepared={
            draw position=39,
            median=1.0,
            upper quartile=1.5,
            lower quartile=-1.5,
            upper whisker=2,
            lower whisker=-3
            },
            ] coordinates {};

            \addplot+[
            boxplot prepared={
            draw position=41,
            median=-2.0,
            upper quartile=-2.0,
            lower quartile=-3.0,
            upper whisker=1,
            lower whisker=-3
            },
            ] coordinates {};

            \addplot+[
            boxplot prepared={
            draw position=42,
            median=-2.0,
            upper quartile=-1.25,
            lower quartile=-2.75,
            upper whisker=3,
            lower whisker=-3
            },
            ] coordinates {};

            \addplot+[
            boxplot prepared={
            draw position=43,
            median=-2.0,
            upper quartile=-1.5,
            lower quartile=-3.0,
            upper whisker=-1,
            lower whisker=-3
            },
            ] coordinates {};

            \addplot+[
            boxplot prepared={
            draw position=45,
            median=-1.0,
            upper quartile=1.0,
            lower quartile=-2.0,
            upper whisker=3,
            lower whisker=-3
            },
            ] coordinates {};

            \addplot+[
            boxplot prepared={
            draw position=46,
            median=-1.0,
            upper quartile=1.0,
            lower quartile=-2.0,
            upper whisker=3,
            lower whisker=-2
            },
            ] coordinates {};

            \addplot+[
            boxplot prepared={
            draw position=47,
            median=0.0,
            upper quartile=1.0,
            lower quartile=-1.75,
            upper whisker=3,
            lower whisker=-3
            },
            ] coordinates {};

            \addplot+[
            boxplot prepared={
            draw position=49,
            median=1.0,
            upper quartile=2.0,
            lower quartile=-1.0,
            upper whisker=3,
            lower whisker=-3
            },
            ] coordinates {};

            \addplot+[
            boxplot prepared={
            draw position=50,
            median=-1.0,
            upper quartile=1.0,
            lower quartile=-2.0,
            upper whisker=3,
            lower whisker=-2
            },
            ] coordinates {};

            \addplot+[
            boxplot prepared={
            draw position=51,
            median=-1.5,
            upper quartile=1.0,
            lower quartile=-2.0,
            upper whisker=3,
            lower whisker=-3
            },
            ] coordinates {};

            \addplot+[
            boxplot prepared={
            draw position=53,
            median=1.0,
            upper quartile=2.0,
            lower quartile=-1.0,
            upper whisker=2,
            lower whisker=-2
            },
            ] coordinates {};

            \addplot+[
            boxplot prepared={
            draw position=54,
            median=1.0,
            upper quartile=1.5,
            lower quartile=-1.0,
            upper whisker=3,
            lower whisker=-2
            },
            ] coordinates {};

            \addplot+[
            boxplot prepared={
            draw position=55,
            median=1.0,
            upper quartile=1.75,
            lower quartile=1.0,
            upper whisker=2,
            lower whisker=-2
            },
            ] coordinates {};

            \addplot+[
            boxplot prepared={
            draw position=57,
            median=0.0,
            upper quartile=1.0,
            lower quartile=-1.0,
            upper whisker=2,
            lower whisker=-2
            },
            ] coordinates {};

            \addplot+[
            boxplot prepared={
            draw position=58,
            median=1.0,
            upper quartile=2.0,
            lower quartile=-1.0,
            upper whisker=3,
            lower whisker=-3
            },
            ] coordinates {};

            \addplot+[
            boxplot prepared={
            draw position=59,
            median=-1.0,
            upper quartile=1.5,
            lower quartile=-2.0,
            upper whisker=2,
            lower whisker=-2
            },
            ] coordinates {};

            \addplot+[
            boxplot prepared={
            draw position=61,
            median=1.0,
            upper quartile=2.0,
            lower quartile=-1.0,
            upper whisker=3,
            lower whisker=-3
            },
            ] coordinates {};

            \addplot+[
            boxplot prepared={
            draw position=62,
            median=-1.0,
            upper quartile=1.0,
            lower quartile=-2.0,
            upper whisker=3,
            lower whisker=-3
            },
            ] coordinates {};

            \addplot+[
            boxplot prepared={
            draw position=63,
            median=1.0,
            upper quartile=2.0,
            lower quartile=-1.0,
            upper whisker=3,
            lower whisker=-3
            },
            ] coordinates {};

            \addplot+[
            boxplot prepared={
            draw position=65,
            median=-1.0,
            upper quartile=1.0,
            lower quartile=-1.0,
            upper whisker=2,
            lower whisker=-2
            },
            ] coordinates {};

            \addplot+[
            boxplot prepared={
            draw position=66,
            median=-1.0,
            upper quartile=1.0,
            lower quartile=-2.0,
            upper whisker=3,
            lower whisker=-2
            },
            ] coordinates {};

            \addplot+[
            boxplot prepared={
            draw position=67,
            median=1.0,
            upper quartile=1.5,
            lower quartile=-2.0,
            upper whisker=2,
            lower whisker=-3
            },
            ] coordinates {};

        \end{axis}
        \end{tikzpicture}
    \end{adjustbox}
    \caption{Boxplots visualizing the Likert scale values evaluated in the first user survey for the outputs from~\Cref{fig:appendix:survey1:outputs}.}
    \label{fig:appendix:survey1:boxplots}
\end{figure}
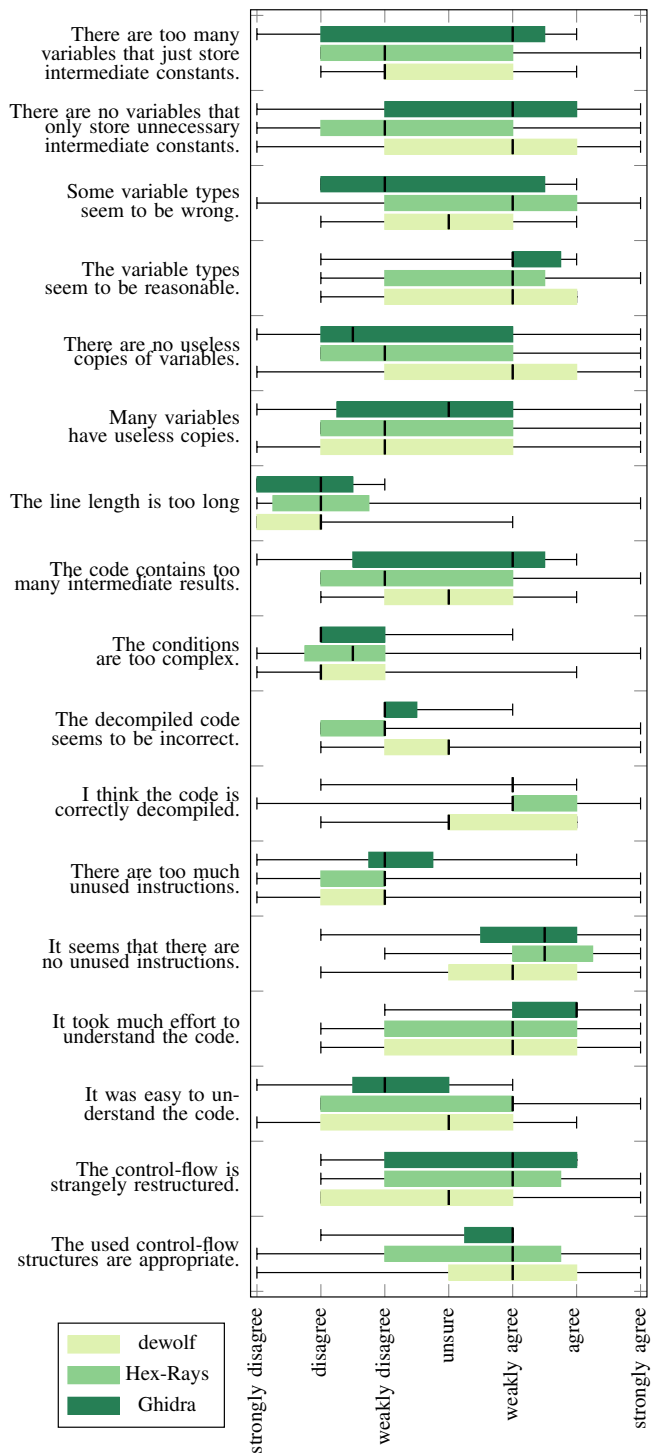

\subsection{User Survey II}\label{appendix:survey2}
To verify the comprehensibility of dewolf's output, we decompiled the DGA of~\Cref{fig:eval:dga} with dewolf.
The output is shown in~\Cref{fig:eval:dga:dewolf}.
\begin{figure}[tb!]
    \begin{minted}[autogobble, bgcolor=my-bg,linenos,numbersep=4pt, breaklines,fontsize=\scriptsize, escapeinside=!!]{c}
        int sub_401000() {
          char i; char * var_0; char var_1; char var_3; int var_4;
          var_0 = malloc(16);
          var_1 = GetTickCount();
          GetSystemTime(lpSystemTime: var_0);
          for (i = 0; i < 8; i++) {
            var_3 = var_0[i];!\label{appendix:line:propagate}!
            var_0[i] = ((unsigned int)(var_3^var_1)%24 &0xff)+'a';!\label{appendix:line:ci1}!
          }
          var_4 = var_0 + 8;
          *var_4 = 0x6d6f632e;
          *(var_4 + 4) = 0x0;
          switch((((int) var_1) % 0x8) - 0x1) { !\label{appendix:line:ci2}!
          case 0:
          case 5:
            *var_4 = *var_4 ^ 0x6d001700;
            break;
          case 1:
            *var_4 = *var_4 ^ 0x190a0d00;
            break;
          case 4:
          case 6:
            *var_4 = *var_4 ^ 0x6d1a1100;
            break;
          }
          return var_0;
        }
    \end{minted}
    \caption{The output of dewolf given the DGA of~\Cref{fig:eval:dga}.}\label{fig:eval:dga:dewolf}
\end{figure}
On the positive side, dewolf correctly restructures the switch statement as well as the array-access for the domains (variable \mintinline{c}{var_0}).
Additionally, dewolf is capable of reversing the compiler idioms in lines~\ref{appendix:line:ci1} \&~\ref{appendix:line:ci2}, e.g., the modulo~24 computation.
Nevertheless, there is still room for improvement.
For example, dewolf does not detect the array access for the TLDs (variable \mintinline{c}{var_4}).
Moreover, dewolf is very conservative with propagation pointers leading to unnecessary copies of variables, for example, in line~\ref{appendix:line:propagate}.
However, we already attacked this problem.
The current version of dewolf propagates the instruction in line~\ref{appendix:line:propagate} into line~\ref{appendix:line:ci1}.
We asked each participant the following comprehension questions:
\begin{itemize}
    \item What does this function return?
        \begin{itemize}
        \item[\single] No Idea
        \item[\single] Manipulates the system time
        \item[\single] Allows Virtual Machine Detection
        \item[\single] Generates random domains
        \item[\single] Encrypts memory regions
        \end{itemize}
    \item Please type all utilized top-level-domains (TLDs).
    \item Which Top-Level domain will be utilized the most?
    \item Which of these second-level domains can potentially be generated by the function? (Yes/No answer option)
      \begin{multicols}{2}
      \begin{itemize}
          \item simpmpfp
          \item xfbcbcic
          \item facebook
          \item squzuzfz
          \item rlpmpmgmjdh
          \item[]
      \end{itemize}
    \end{multicols}
\end{itemize}
The correct answer to the first question is \emph{Generates random domains}.
To identify all utilized top-level domains, one has to look at the strings assigned to the array~\mintinline{c}{end} resp.\ the pointer~\mintinline{c}{var_4}.
This leads to the following four TLDs:
\begin{itemize}
  \item \texttt{.com} \hspace{5mm}(\mintinline[fontsize=\small]{c}{"\x2e\x63\x6f\x6d"})
  \item \texttt{.ru }  \hspace{5mm}(\mintinline[fontsize=\small]{c}{"\x00\x11\x1a\x6d"})
  \item \texttt{.tu }  \hspace{5mm}(\mintinline[fontsize=\small]{c}{"\x00\x17\x00\x6d"})
  \item \texttt{.net} \hspace{5mm}(\mintinline[fontsize=\small]{c}{"\x00\x0d\x0a\x19"}).
\end{itemize}
The most utilized TLD is \texttt{.com} because we have eight different cases and \texttt{.com} is used for three, whereas the others are only chosen for up to two cases.
Finally, the only second-level domains not potentially generated are \emph{squzuzfz} and \emph{rlpmpmgmjdh} because the character z is excluded in the for-loop and the length of each domain is eight.

\subsection{User Survey III}
During the first part of the third survey, each participant was given and asked to rank all three decompiler outputs from~\Cref{fig:appendix:survey3:outputs}.
The ranking results are illustrated in~\Cref{fig:eval:ranking} and shortly discussed in~\Cref{sec:evaluation}.
\Cref{fig:appendix:survey3:hexrays} shows that Hex-Rays restructured the loop-body very closely to the structure of the assembly code by utilizing an if/else structure.
Similarly, Ghidra also produced an if/else structure, but with a significantly higher nesting depth by utilizing cascading if statements (\Cref{fig:appendix:survey3:ghidra}).
In contrast to both, dewolf restructured this part into a switch statement by identifying a candidate variable for a switch construct (\Cref{fig:appendix:survey3:dewolf}).
Although this does not strictly reflect the assembly due to the missing jump table calculation, we argue it is easier to understand.
Further differences between the outputs mentioned by the participants include but are not limited to:
\begin{itemize}
  \item The loop-type used for restructuring, i.e., Ghidra and dewolf output a for-loop and Hex-Rays a while-loop.
  \item The mixed constant representation of dewolf.
  \item The usage of casts, i.e., Ghidra used nearly twice as many casts than IDA and that dewolf used no casts.
  \item Hex-Rays superior type and array reconstruction.
\end{itemize}

\begin{figure*}[p!]
  \begin{subfigure}[l]{0.32\linewidth}
    \begin{minted}[autogobble, bgcolor=my-bg,linenos,numbersep=4pt, breaklines,fontsize=\fontsize{7}{7.4}, fontfamily={tt}, fontseries={console-like}]{c}
    undefined8 FUN_00401c3d(void){
      size_t sVar1; undefined8 local_68;
    \end{minted}
    \vspace{-0.9cm}
    \begin{minted}[bgcolor=my-bg,linenos,numbersep=4pt,escapeinside=||,breaklines,fontsize=\fontsize{7}{7.4}]{c}
|\setcounter{FancyVerbLine}{12}|  long local_18; int local_10;
  uint local_c; local_68 = 0;
    \end{minted}
    \vspace{-0.9cm}
    \begin{minted}[bgcolor=my-bg,linenos,numbersep=4pt,escapeinside=||,breaklines,fontsize=\fontsize{7}{7.4}]{c}
|\setcounter{FancyVerbLine}{24}|  printf("Enter any binary number: ");
  __isoc99_scanf(&DAT_00403259, &local_18);
  local_c = (uint)local_18;
  for (; 0 < local_18; local_18 = local_18 / 10000) {
  local_10 = (int)local_18 + (int)(local_18 / 10000) *-10000;
  if (local_10 == 0x457) {
    sVar1 = strlen((char *)&local_68);
    *(undefined2 *)((long)&local_68 + sVar1) = 0x46; }
  else { if (local_10 < 0x458) {
    if (local_10 == 0x456) {
      sVar1 = strlen((char *)&local_68);
      *(undefined2 *)((long)&local_68 + sVar1) = 0x45; }
    else { if (local_10 < 0x457) {
      if (local_10 == 0x44d) {
        sVar1 = strlen((char *)&local_68);
        *(undefined2 *)((long) &local_68 + sVar1) = 0x44; }
      else { if (local_10 < 0x44e) {
        if (local_10 == 0x44c) {
    \end{minted}
    \vspace{-0.9cm}
    \begin{minted}[bgcolor=my-bg,linenos,numbersep=4pt,escapeinside=||,breaklines,fontsize=\fontsize{7}{7.4}]{c}
|\setcounter{FancyVerbLine}{105}|  printf("Binary number: %lld\\n",(ulong)local_c);
  printf("Hexadecimal number: %s",&local_68);
  return 0; }
      \end{minted}
      \subcaption{Output generated with the decompiler integrated into Ghidra~10.0.3.}
      \label{fig:appendix:survey3:ghidra}
  \end{subfigure}\hfill
    \begin{subfigure}[l]{0.32\linewidth}
        \begin{minted}[bgcolor=my-bg,linenos,numbersep=4pt,escapeinside=||,breaklines,fontsize=\fontsize{7}{7.4}]{c}
long sub_401c3d() {
  size_t var_5; long i;
  long var_0; long var_2; long var_4;
  long * var_3;
  printf(/* format */ "Enter any binary number: ");
  var_3 = &var_0;
  __isoc99_scanf(/* format */ "%lld", var_3);
  var_2 = 0L;
  for(i = var_0; i > 0L; i /= 0x2710){
    var_4 = i % 0x2710;
    switch(var_4) {
    case 0:
      var_3 = &var_2;
      var_5 = strlen(var_3);
      *(&var_2 + var_5) = 0x30; break;
    case 1:
      var_3 = &var_2;
      var_5 = strlen(var_3);
      *(&var_2 + var_5) = 0x31; break;
    case 10:
      var_3 = &var_2;
      var_5 = strlen(var_3);
      *(&var_2 + var_5) = 0x32; break;
    case 11:
      var_3 = &var_2;
      var_5 = strlen(var_3);
      *(&var_2 + var_5) = 0x33; break;
        \end{minted}
        \vspace{-0.9cm}
        \begin{minted}[bgcolor=my-bg,linenos,numbersep=4pt,escapeinside=||,breaklines,fontsize=\fontsize{7}{7.4}]{c}
  |\setcounter{FancyVerbLine}{76}| printf(/* format */ "Binary number: %lld\\n", var_0 & 0xffffffff);
  var_3 = &var_2;
  printf(/* format */ "Hexadecimal number: %s", var_3);
  return 0L;
        \end{minted}
        \vspace{0.45cm}
        \subcaption{Output generated with dewolf integration all improvements from~\Cref{sec:approach}.}
        \label{fig:appendix:survey3:dewolf}
    \end{subfigure}\hfill
    \begin{subfigure}[l]{0.32\linewidth}
        \begin{minted}[autogobble, bgcolor=my-bg,linenos,numbersep=4pt,escapeinside=||,breaklines,fontsize=\fontsize{7}{7.4}]{c}
        _int64 sub_401C3D(){
          char s[8]; // [rsp+0h] [rbp-60h] BYREF
          __int64 v2; // [rsp+8h] [rbp-58h]
        \end{minted}
        \vspace{-0.9cm}
        \begin{minted}[bgcolor=my-bg,linenos,numbersep=4pt,escapeinside=||,breaklines,fontsize=\fontsize{7}{7.4}]{c}
  |\setcounter{FancyVerbLine}{9}|__int64 v8; // [rsp+38h] [rbp-28h]
  char v9; // [rsp+40h] [rbp-20h]
  __int64 v10; // [rsp+50h] [rbp-10h] BYREF
  int v11; // [rsp+58h] [rbp-8h]
  unsigned int v12; // [rsp+5Ch] [rbp-4h]
  *(_QWORD *)s = 0LL;
  v2 = 0LL;
        \end{minted}
        \vspace{-0.9cm}
        \begin{minted}[bgcolor=my-bg,linenos,numbersep=4pt,escapeinside=||,breaklines,fontsize=\fontsize{7}{7.4}]{c}
|\setcounter{FancyVerbLine}{23}|  v9 = 0;
  printf("Enter any binary number: ");
  __isoc99_scanf("%lld", &v10); v12 = v10;
  while ( v10 > 0 ) {
  v11 = v10 % 10000;
  if ( v11 == 1111 ) {
    *(_WORD *)&s[strlen(s)] = 70; }
  else if ( v11 <= 1111 ) {
    if ( v11 == 1110 ) {
      *(_WORD *)&s[strlen(s)] = 69; }
    else if ( v11 == 1101 ) {
      *(_WORD *)&s[strlen(s)] = 68; }
    else if ( v11 <= 1101 ) {
      if ( v11 == 1100 ) {
        *(_WORD *)&s[strlen(s)] = 67; }
          \end{minted}
          \vspace{-0.9cm}
          \begin{minted}[bgcolor=my-bg,linenos,numbersep=4pt,escapeinside=||,breaklines,fontsize=\fontsize{7}{7.4}]{c}
  |\setcounter{FancyVerbLine}{68}|v10 /= 10000LL; }
  printf("Binary number: %lld\\n", v12);
  printf("Hexadecimal number: %s", s);
  return 0LL;}
        \end{minted}
        \vspace{-0.1cm}
        \subcaption{Output generated with Hex-Rays and IDA 7.61 (SP1).}
        \label{fig:appendix:survey3:hexrays}
    \end{subfigure}\hfill
    \caption{Decompiler output excerpts for the function from~\Cref{fig:eval:sample_survey3} used for the third user survey.}
    \label{fig:appendix:survey3:outputs}
\end{figure*}

In the second part, we questioned the general assumption that the structure of the decompiled output should be similar to the assembly structure.
Therefore, we restructured a sample in three different ways, once by copying a sequence of code (\Cref{appendix:fig:copy}), one by using an additional variable to reconstruct the flow (\Cref{appendix:fig:exit}), and once by extracting part of the code into a function (\Cref{appendix:fig:function}).
Additionally to the comparison in~\Cref{appendix:fig:extraction}, we also asked the same question when the marked region consists of one, five, or~15 lines of code.
As expected, copying one line of code was alright when it simplifies the structure.
However, the more complicated and longer the structure gets, the participants do not like that it is copied.
The idea of extracting complicated structures into a function was picked up very positively by most participants.
  \begin{figure*}[p!]
      \begin{subfigure}[t]{0.32\textwidth}
          \begin{minted}[autogobble, bgcolor=my-bg,linenos, breaklines,fontsize=\fontsize{7}{8},numbersep=4pt, highlightlines={9-15, 20-26},highlightcolor=my-bg!90!black]{c}
              /* Block #0 */
              if(var_0 > 10){
                while(var_1 > 0){
                  if(var_2 == 2){ /* Block #4 */
                    if(var_4 == 4){ /* Block #8 */
                      continue;}
                    /* Block #9 */
                  }else{/* Block #5 */}
                  while(var > 1){
                    printf("Enter a number \\n");
                    scanf("%d", &numb1);
                    if(var % numb == 0){var/=numb;}
                    else{var -= numb;}
                    printf("The new number is %d, \\n", var);}
                  return var;}}
              /* Block #3 */
              if(var_3 > 3){/* Block #6 */}
              else{/* Block #7 */}
              /* Block #11 */
              while(var > 1){
                printf("Enter a number \\n");
                scanf("%d", &numb1);
                if(var % numb == 0){var /= numb;}
                else{var -= numb;}
                printf("The new number is %d, \\n", var);}
              return var;
          \end{minted}
          \subcaption{Copying the marked region.}
          \label{appendix:fig:copy}
      \end{subfigure}\hfill
      \begin{subfigure}[t]{0.32\textwidth}
          \begin{minted}[autogobble, bgcolor=my-bg,linenos, breaklines,fontsize=\fontsize{7}{8},numbersep=4pt, highlightlines={18-24},highlightcolor=my-bg!90!black]{c}
              /* Block #0 */
              if(var_0 > 10){
                while(true){
                  if(var_1 > 0){
                    exit_1 = 0;
                    break;}
                  if(var_2 == 2){ /* Block #4 */
                    if(var_4 == 4){ /* Block #8 */
                      continue;}
                    /* Block #9 */
                  }else{/* Block #5 */}
                  exit = 1;
                  break;}}
              if(var_0 <= 10 || exit_1 == 0){
                if(var_3 > 3){/* Block #6 */}
                else{/* Block #7 */}
                /* Block #11 */}
              while(var > 1){
                printf("Enter a number \\n");
                scanf("%d", &numb1);
                if(var % numb == 0){var /= numb;}
                else{ var -= numb; }
                printf("The new number is %d, \\n", var); }
              return var;
          \end{minted}
          \vspace{1.19em}
          \subcaption{Adding an additional variable~\mintinline{c}{exit_1}.}
          \label{appendix:fig:exit}
      \end{subfigure}\hfill
      \begin{subfigure}[t]{0.32\textwidth}
          \begin{minted}[autogobble, bgcolor=my-bg,linenos, breaklines,fontsize=\fontsize{7}{8},numbersep=4pt, highlightlines={17-23, 9, 14},highlightcolor=my-bg!90!black]{c}
              /* Block #0 */
              if(var_0 > 10){
                while(var_1 > 0){
                  if(var_2 == 2){ /* Block #4 */
                    if(var_4 == 4){ /* Block #8 */
                      continue;}
                    /* Block #9 */
                  }else{/* Block #5 */}
                  return call_sub(var);}}
              /* Block #3 */
              if(var_3 > 3){/* Block #6 */}
              else{/* Block #7 */}
              /* Block #11 */
              return call_sub(var);

              int call_sub(int var){
                while(var > 1){
                  printf("Enter a number \\n");
                  scanf("%d", &numb1);
                  if(var % numb == 0){var /= numb;}
                  else{var -= numb;}
                  printf("The new number is %d, \\n", var);}
                return var;}
          \end{minted}
          \vspace{1.96em}
          \subcaption{Extracting part of the code into a function.}
          \label{appendix:fig:function}
      \end{subfigure}
      \caption{The comparison of different restructuring options to avoid gotos for loops with multiple exits.}
      \label{appendix:fig:extraction}
  \end{figure*}

\end{document}